\documentclass[10pt,conference,compsocconf]{IEEEtran}

\usepackage[nocompress]{cite}
\usepackage[pdftex]{graphicx}
\graphicspath{{figs/}}
\DeclareGraphicsExtensions{.pdf,.svd, .jpeg,.png}

\makeatletter
\renewcommand{\@IEEEsectpunct}{\ \,}
\makeatother

\usepackage{amsmath}
\usepackage[dvipsnames]{xcolor}
\usepackage[normalem]{ulem}
\usepackage{comment}
\usepackage{booktabs} 
\usepackage{kotex}
\usepackage{enumitem}
\usepackage{subcaption}
\usepackage[font=footnotesize]{caption}
\usepackage{multirow}
\usepackage{blindtext}
\usepackage{tabulary}
\usepackage{listings}
\usepackage{makecell}
\usepackage{textcomp}
\usepackage[export]{adjustbox}
\usepackage{soul}

\definecolor{applegreen}{rgb}{0.55, 0.71, 0.0}

\newcommand{\return}{\textbf{return} }

\usepackage{pifont}
\newcommand{\bcheckmark}{\ding{51}} 
\newcommand{\btimes}{\ding{55}}     

\usepackage{algorithmic}
\usepackage[linesnumbered,ruled,vlined]{algorithm2e}

\SetCommentSty{mycommfont}
\SetAlFnt{\footnotesize}
\SetAlCapFnt{\footnotesize}
\SetAlCapNameFnt{\footnotesize}

\usepackage{draftwatermark}
\SetWatermarkText{Preprint}
\SetWatermarkScale{0.5}
\SetWatermarkColor[gray]{0.9}

\usepackage{array}
\usepackage{hyperref}
\hypersetup{
	colorlinks=true,
	linkcolor=blue,
	filecolor=blue,      
	urlcolor=blue,
	pdfpagemode=FullScreen
}

\usepackage{tikz}
\usepackage{pgfplots}
\pgfplotsset{compat=1.7}
\usetikzlibrary{math,positioning,calc}
\usetikzlibrary{pgfplots.statistics,pgfplots.colorbrewer}
\usetikzlibrary{external}
\tikzset{external/system call={pdflatex \tikzexternalcheckshellescape -halt-on-error -interaction=batchmode -jobname "\image" "\texsource"}}
\tikzsetexternalprefix{figures-cached/}
\tikzset{external/only named=true}
\newcommand*\circled[1]{\tikz[baseline=(char.base)]{
            \node[shape=circle,draw,inner sep=0.8pt] (char) {{\textbf{#1}}};}}

\newcommand{\systemname}{\textsc{OctopInf}}
\newcommand{\cwd}{\textsc{Cwd}}
\newcommand{\coral}{\textsc{Coral}}

\newcommand{\sourcecodelink}{https://github.com/tungngreen/PipelineScheduler}
\newcommand{\earthcam}{www.earthcam.com}

\begin{document}

\setlength{\abovedisplayskip}{1pt}
\setlength{\belowdisplayskip}{1pt}
\setlength{\textfloatsep}{0.1cm}
\setlength{\floatsep}{0.1cm}

\title{\systemname{}: Workload-Aware Inference Serving for Edge Video Analytics}

\author{\IEEEauthorblockN{Thanh-Tung Nguyen, Lucas Liebe, Nhat-Quang Tau, Yuheng Wu, Jinghan Cheng, Dongman Lee}
\IEEEauthorblockA{
School of Computing, KAIST, Republic of Korea\\
\{tungnt, lucasliebe, quangntau1223, yuhengwu, chengjh, dlee\}@kaist.ac.kr (\textit{Correspondence}: \{tungnt, dlee\}@kaist.ac.kr)}} 

\maketitle

\begingroup\renewcommand\thefootnote{\IEEEauthorrefmark{1}}
\footnotetext{This paper has been accepted to IEEE PerCom 2025. © 2025 IEEE. The final version will be available at [DOI link when available].
}


\begin{abstract}
Edge Video Analytics (EVA) has become a major application of pervasive computing, enabling real-time visual processing.
EVA pipelines, composed of deep neural networks (DNNs), typically demand efficient inference serving under stringent latency requirements, which is challenging due to the dynamic Edge environments (e.g., workload variability and network instability).
Moreover, EVA pipelines face significant resource contention due to resource (e.g., GPU) constraints at the Edge.
In this paper, we introduce \systemname{}, a novel resource-efficient and workload-aware inference serving system designed for real-time EVA.
\systemname{} tackles the unique challenges of dynamic edge environments through fine-grained resource allocation, adaptive batching, and workload balancing between edge devices and servers. Furthermore, we propose a spatiotemporal scheduling algorithm that optimizes the co-location of inference tasks on GPUs, improving performance and ensuring service-level
objectives (SLOs) compliance. Extensive evaluations on a real-world testbed demonstrate the effectiveness of our approach. It achieves an effective throughput increase of up to $10\times$ compared to the baselines and shows better robustness in challenging scenarios. \systemname{} can be used for any DNN-based EVA inference task with minimal adaptation and is available at \href{\sourcecodelink}{\sourcecodelink}.
\end{abstract}

\IEEEpeerreviewmaketitle

\section{Introduction}
\label{sec:intro}

\par
Recently, \textit{Edge Video Analytics (EVA)} has emerged as a major area of pervasive computing \cite{bahl2020percom}, offering real-time visual sensing and processing.
The pervasive nature of EVA systems enables seamless integration into diverse tasks such as surveillance~\cite{Pasandi2020convince, Hayashi2022traffic, Nguyen2023preacto, Anjum2024surveillance}, health care \cite{Sahu2023healthcare}, and activity recognition \cite{Kumrai2020activityrecognition, Bicocchi2012activityrecognition}.
These systems can perform continuous, on-site video stream analysis with reduced dependency on remote cloud infrastructures.
In practice, VA services are typically organized as cascading pipelines of deep neural networks (DNNs)\cite{Jang2021pipeline}.
For instance, the pipeline of \linebreak $[$Object Detect $\rightarrow$ Vehicle Classify, Plate Detect$]$  can be employed for traffic monitoring (\autoref{fig:content_dynamics}).
The task of executing the pipeline is defined as \textit{inference serving} and is subjected to stringent latency demands (e.g., 200 ms), specified by service-level objectives (SLOs).
It has been studied in works such as \cite{shen2019nexus, Choi2022Gpulet, crankshaw2020inferline,  nigade2022jellyfish, zeng2020distream, Hou2023dystri} with the common goal of efficiently allocating and scheduling resources to meet SLOs.

\par
Advancements in embedded computing now enable DNN models to run on both Edge servers and nearby embedded devices (e.g., Jetson Orin Nano) deployed next to data sources such as CCTV cameras \cite{Mendula2024deviceserversplit}.
Despite their limited capabilities, Edge devices can execute part of the EVA pipeline, enabling server-device cooperation to share workloads \cite{Jang2021pipeline,zeng2020distream, Hou2023dystri} increasing efficiency and flexibility.
However, implementing this deployment scenario to reap its full potential involves two key considerations. \linebreak
(1) \textit{Highly dynamic environments}. At the Edge, significant variations in video content over time lead to changes in workloads for models within an EVA pipeline (\autoref{fig:content_dynamics}), and constant changes in network conditions (i.e., bandwidth and latency) cause fluctuations in compute time budgets for the pipeline, complicating the workload distribution task. 
\linebreak
(2) \textit{Resource contention}. At the Edge, multiple DNN models run concurrently, or \textit{co-located}, on the same processing unit (e.g., GPUs), leading to unpredictable performance degradations in terms of latency (defined as \textit{co-location interference} \cite{Wu2022colocationinference}).
While the adverse effects can be lessened by grouping models with low workloads, the workload dynamicity mentioned above makes this impractical.

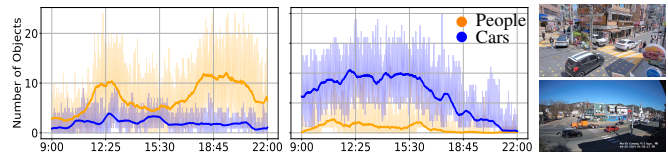
\begin{figure}[t]
    \centering
    \hspace*{-0.35cm} 
    \subfloat{
        \includegraphics[width=0.405\linewidth, height=2.0cm, trim=0.8cm 0cm 2.2cm 1.5cm, clip]{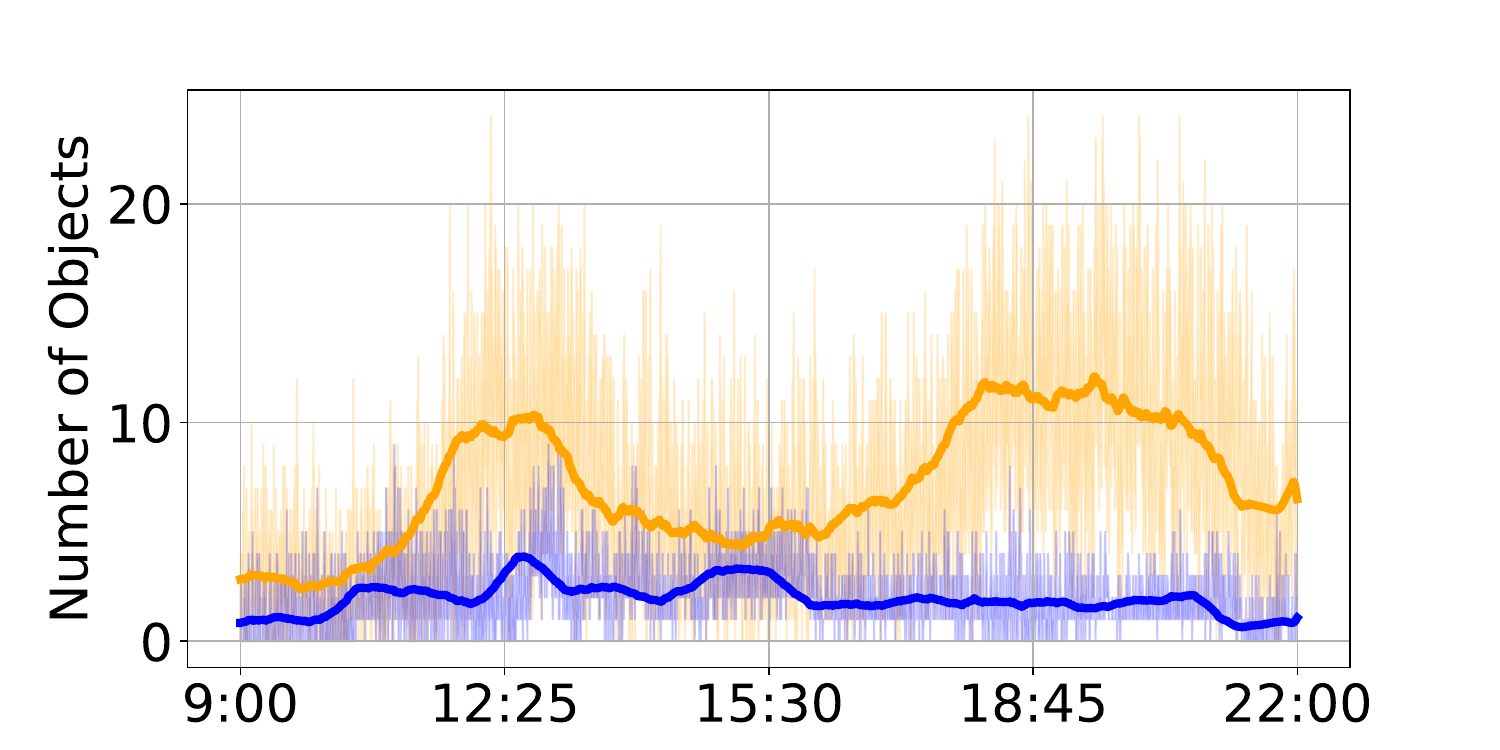}
    }
    \hspace{-8pt}
    \subfloat{
        \begin{tikzpicture}
            \node[anchor=south west, inner sep=0] (image) at (0,0) {\includegraphics[width=0.365\linewidth, height=2.0cm, trim=3cm 0cm 2.2cm 1.5cm, clip]{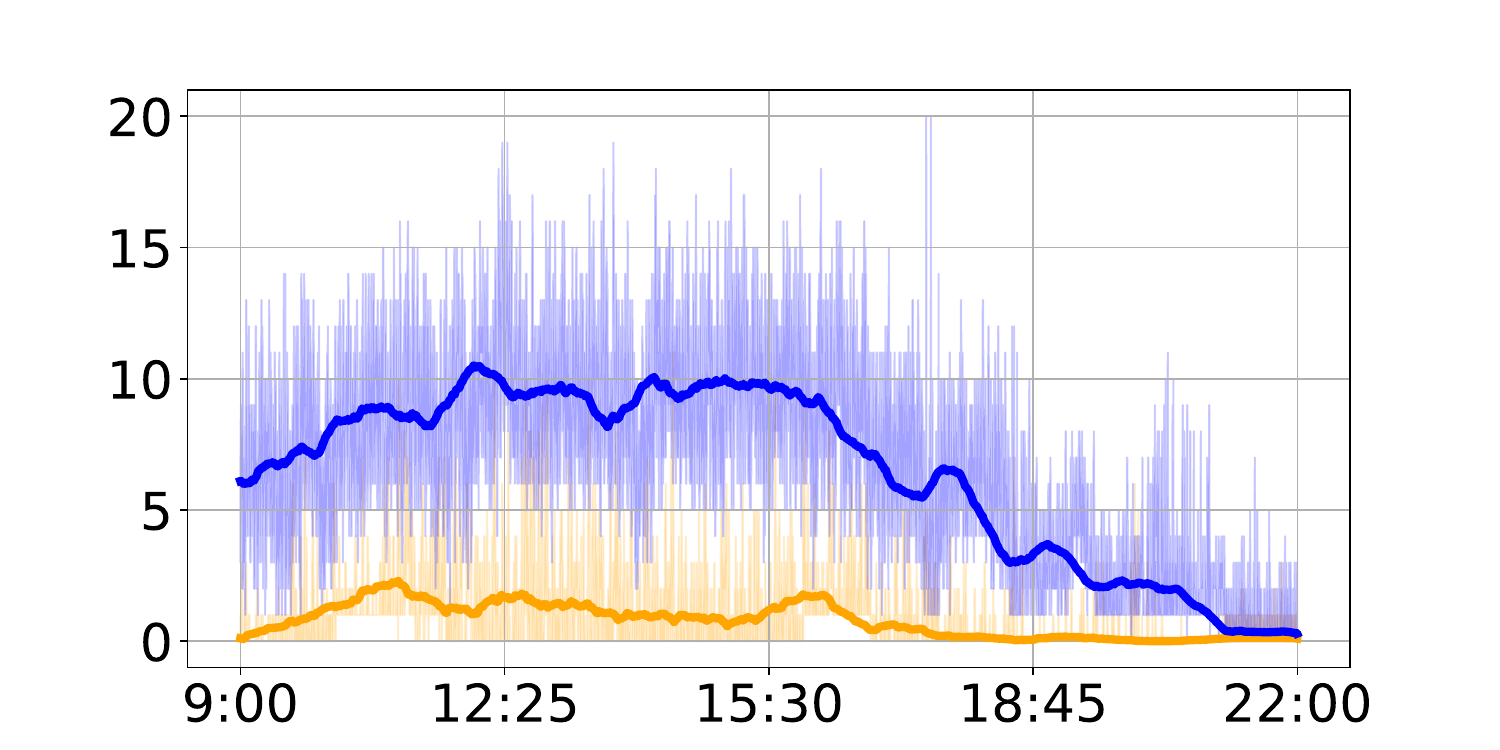}};
            \begin{scope}[x={(image.south east)}, y={(image.north west)}]
                \fill[orange] (0.71,0.9) circle (2pt);
                \node[right] at (0.73,0.9) {\rotatebox{0}{\textnormal{\scriptsize People}}};
                \fill[blue] (0.71,0.8) circle (2pt);
                \node[right] at (0.73,0.8) {\rotatebox{0}{\textnormal{\scriptsize Cars}}};
            \end{scope}
        \end{tikzpicture}
    }
    \hspace{-9pt}
    \subfloat{     
        \vspace{0.08cm}
        \begin{minipage}[b]{0.19\linewidth}
            \centering
            \includegraphics[width=\linewidth]{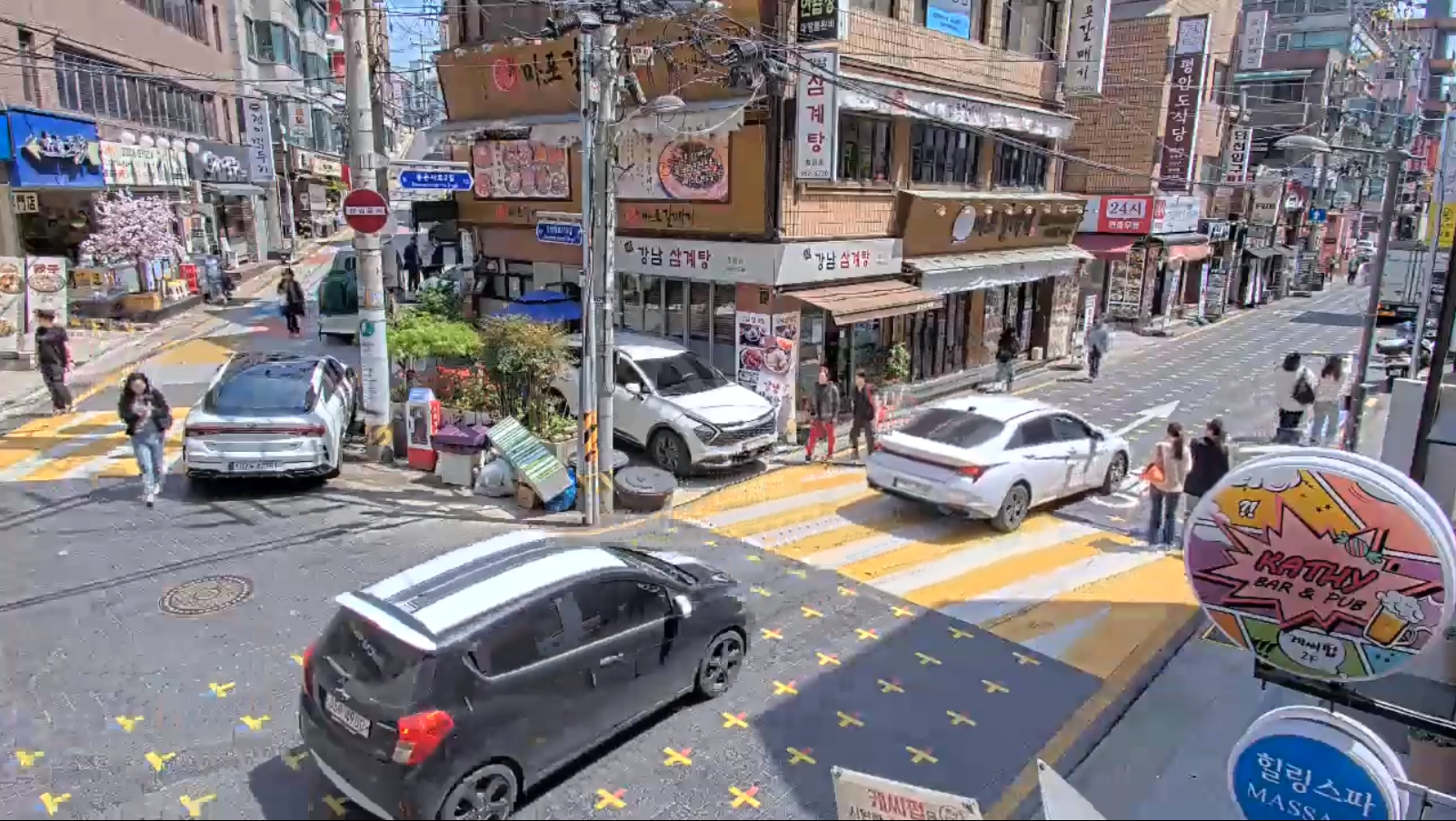}\\[0.15em]
            \includegraphics[width=\linewidth]{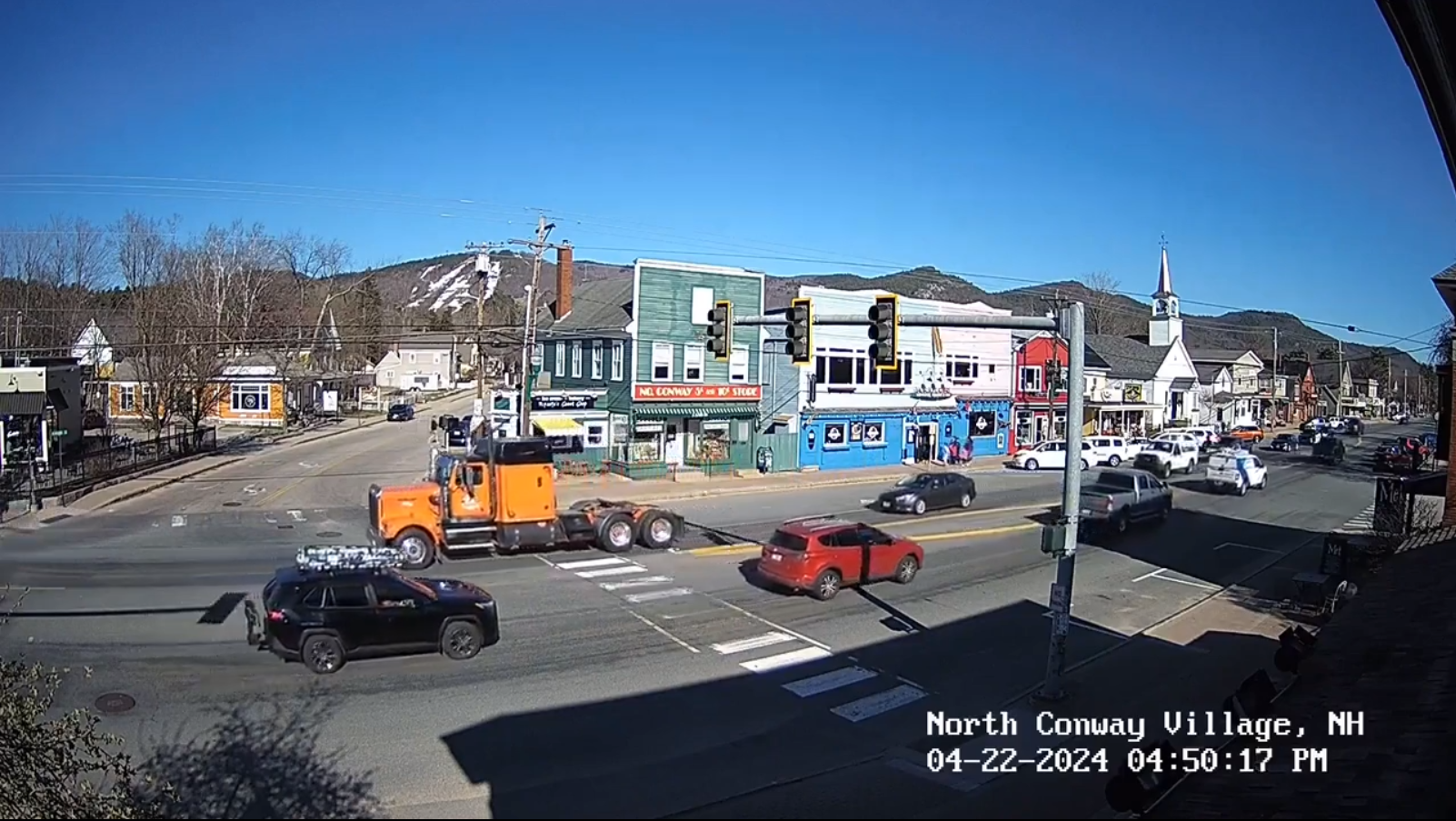}
        \end{minipage}
    }
    \vspace{-0.14cm}
    \caption{Real-world footage at 2 locations shows varying number of objects leading to variations in workload for pipeline models (e.g., car classification).}
    \label{fig:content_dynamics}
\end{figure}

\par
Several solutions have been proposed. Jellyfish \cite{nigade2022jellyfish} leverages \textit{dynamic batching} to increase throughput and uses multiple DNN versions to adapt to network latency variations.
During network instability, it reduces data resolution by opting for less accurate model versions, and adjusts batch sizes to meet throughput demands.
However, its centralized architecture, which transfers raw videos to the server and assumes ample GPU resources to avoid co-location interference, is unsuitable for resource-constrained Edge environments.
Furthermore, Jellyfish can only adjust batch sizes for different versions of the same model, not systematically schedule the whole pipeline.
Distributed architectures like Distream \cite{zeng2020distream} and Rim \cite{hu2021rim} instead utilize Edge devices and balance workloads between devices and servers. Distream introduces a stochastic method to determine the "\textit{split point}," adaptively dividing EVA pipelines between local and server-based workloads.
However, to reduce the optimization space, it uses a static batch size for models, which fails to account for dynamic workloads and network conditions.
This is because while batching improves throughput, it increases end-to-end latency for queries in the batch, risking SLO violations \cite{crankshaw2020inferline}. 
Batch sizes must therefore be dynamically adjusted to real-time workloads and conditions.
Rim \cite{hu2021rim} argues that Edge models rarely benefit from batching due to lower workloads compared to the cloud.
It selects the \textit{split point} by maximizing concurrent model execution to improve hardware utilization.
However, this approach does not hold under dynamic Edge conditions.
For example, during rush-hour traffic, a car-type classifier may experience high query volumes that benefit significantly from batched inference.
Otherwise, without batching, throughput will suffer. Furthermore, all distributed architectures place multiple inference models on the same GPU without addressing the performance degradations caused by \textit{co-location interference} \cite{Wu2022colocationinference}.
In short, these works can only solve a small part of the problem set, making their solutions incomplete.

\begin{table}[!t]
    \centering
    \caption{Comparisons to the state-of-the-art EVA inference systems}
    \footnotesize
    \vspace{-0.7em}
    \label{Tab:sota-comparison}
    \begin{tabular}{@{\hspace{1pt}} >{\centering\arraybackslash}p{1.3cm} 
                    >{\centering\arraybackslash}p{1.6cm}
                    @{\hspace{2pt}} >{\centering\arraybackslash}p{1.4cm}
                    @{\hspace{5pt}} >{\centering\arraybackslash}p{2.2cm}
                    @{\hspace{2pt}} >{\centering\arraybackslash}p{1.3cm}}
                    
      \hline
      \toprule
      \textbf{System} &
      \makecell{\textbf{Workload} \\ \textbf{Distribution}} & 
      \makecell{\textbf{Dynamic} \\ \textbf{Batching}} &
      \makecell{\textbf{Spatiotemporal} \\ \textbf{GPU Scheduling}} &
      \makecell{\textbf{Horizontal} \\ \textbf{Scaling}} \\\hline

      Jellyfish~\cite{nigade2022jellyfish} & \textit{centralized} & \textit{single tasks} & \btimes & \btimes \\\hline
      Distream~\cite{zeng2020distream} & \bcheckmark & \btimes & \btimes & \btimes \\\hline
      Rim~\cite{hu2021rim} & \bcheckmark & \btimes & \btimes   & \btimes \\\hline
      \textbf{\systemname{}} & \bcheckmark & \textit{pipeline} & \bcheckmark & \bcheckmark \\\hline
    \end{tabular}
    \vspace{0.5em}
\end{table}

\par
\textbf{Goal and Insight.}
We aim to propose a system that tackles the complete problem of Edge inference serving, addressing limitations of the SOTAs, improving throughput, and reducing latency..
Our key insight is that (1) systematically adjusting model placements and dynamically configuring batch sizes can eliminate computation and communication bottlenecks, enhancing robustness to environmental dynamicity. Additionally, (2) temporally multiplexing model executions mitigates \textit{co-location interference}, improving resource efficiency while ensuring compliance with SLO and throughput demands.

\par
\textbf{Technical Challenges.} Inference serving at the Edge is a highly complex Integer Linear Program (\autoref{sec:problemstatement}) and poses two non-trivial challenges.

\textbf{\textit{1) How can workloads be efficiently distributed while determining optimal batch sizes for pipeline models?}}
Network and content dynamics significantly affect request rates, which vary widely among models.
During high workload periods, increasing a model's batch size boosts throughput, raising request rates for downstream neighbors.
Yet, batch selection for multiple models must occur simultaneously, creating a vast optimization space.
Additionally, while larger batches enhance throughput, they also add latency, requiring a careful balance based on workload dynamics and latency constraints.
Moreover, executing the entire pipeline on the server is inefficient, as transmitting raw data over unstable networks adds overhead and delay.
Collaborative execution between servers and devices, where only essential information is transmitted, mitigates these issues but adds complexity.
\textit{Together, these factors make a highly complex challenge.}

\textbf{\textit{2) How can models be multiplexed to avoid interference while satisfying SLOs and maintaining resource efficiency?}}
Multiplexing model executions on shared GPUs demands a balance between maximizing parallelism and minimizing interference, which requires consideration of spatial aspects such as GPU memory and computation capability.
Scheduling is further complicated by temporal dependencies, as upstream models must process data before the downstreams.
Additionally, 
Current software stacks (e.g., NVIDIA CUDA, Intel OpenVINO) support only kernel-level scheduling, lacking model-level coordination.
\textit{This limitation increases the challenge while creating opportunities for custom strategies to improve resource usage and system performance.}

\par
\textbf{Proposed Approach.}
In this paper, we propose \systemname{}, a workload-aware real-time inference serving system for EVA, designed to tackle the mentioned challenges with 3 main components.
\textbf{First}, the \textbf{Cross-device Workload Distributor} (\textbf{\cwd{}}), pronounced \textit{seaweed}, employs a \textit{workload-aware} greedy algorithm to optimize resource allocation through \textbf{dynamic batching} and balanced workload distribution between Edge devices and the server.
\cwd{} leverages environmental factors like workload burstiness to guide batch size exploration and uses IO ratios to filter out inefficient placement options, minimizing the optimization space for higher throughput and reduced latency.
\textbf{Second}, the \textbf{Co-location Inference Spatiotemporal Scheduler} (\textbf{\coral{}}) coordinates the execution of co-located models.
To enable \coral{}, we introduce the \textit{inference stream} abstraction, simplifying resource allocation and model execution sequencing.
Using a temporally best-fit algorithm with spatial constraints, \coral{} ensures resource usage remains within hardware limits, reducing interference and enhancing efficiency while meeting SLO requirements.
Additionally, \systemname{} incorporates a \textbf{Horizontal Auto Scaler} to adapt to abrupt workload changes, ensuring scalability and robustness in dynamic settings.
Comparisons to SOTAs on various categories are summarized in \autoref{Tab:sota-comparison}.

\par
\systemname{} is designed to integrate seamlessly with various inference platforms (e.g., NVIDIA-TensorRT, Intel-OpenVINO, ONNX) using native programming APIs.
To validate its effectiveness, we implemented \systemname{} on a real-world testbed featuring industry-standard Edge devices and EVA workloads.
Our \textbf{contributions} are as follows:
\begin{itemize} [wide]
    \item  We provide a comprehensive formulation and analysis of the EVA inference optimization problem, addressing the unique challenges posed by dynamic Edge environments and resource constraints at run-time (\autoref{sec:problemstatement}).
    \item We propose \cwd{}, an efficient \textbf{Cross-device Workload Distributor}, and a workload-aware greedy algorithm to adjust batch sizes dynamically and balances workloads, leveraging factors such as workload burstiness and IO ratio, to minimize latency and maximize throughput (\autoref{sec:system-design}).
    \item We introduce \textit{inference stream}, a novel GPU resource abstraction and present \coral{}, a \textbf{Co-location Inference Spatiotemporal Scheduler}, which uses \textit{stream} to simplify resource allocation and enable precise execution sequencing, reducing contention and enhancing throughput (\autoref{sec:system-design}).
    \item We deploy \systemname{} on a real-world testbed with actual Edge video data and show up to $10\times$ \textit{effective throughput} improvement over state-of-the-arts (SOTAs) (\autoref{sec:evaluation}).
\end{itemize}

\section{Problem Formulation}
\label{sec:problemstatement}

\par
We follow a common scenario, extensively considered by both academia~\cite{Jang2021pipeline, zeng2020distream, Hou2023dystri} and the industry~\cite{Ananthanarayanan2020rocket}.
The EVA pipeline consists of multiple DNN models (\autoref{fig:pipelines}), organized as a directed acyclic graph (DAG).
A cluster of \textbf{Edge devices} are connected to real-time data sources and the pipeline workload can be partially offloaded to an \textbf{Edge server} collaboratively~\cite{Hung2018VideoEdge, Nguyen2023preacto}.
The\textbf{ main objective} of~\systemname{} is to maximize the \textbf{\textit{effective throughput}}, defined as \textit{the number of inference results that meet SLOs every second} while also minimizing GPU memory usage and end-to-end latency.

\begin{figure}[t!]
    \begin{center}
    \includegraphics[width=\linewidth,trim=0cm 0cm 0cm 0.1cm, clip]{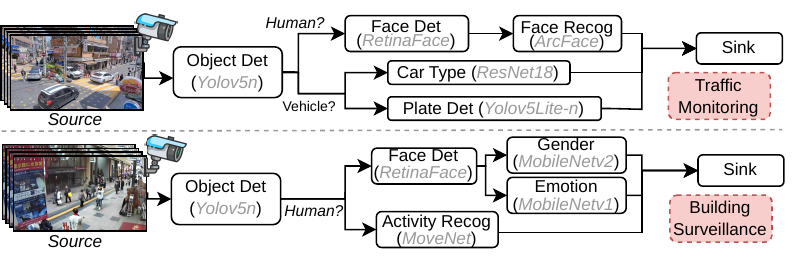} 
    \vspace{-0.7cm}
    \caption{Examples of common EVA pipelines.} 
    \label{fig:pipelines}
    \end{center}
    \vspace{-0.5em}
\end{figure}

\par
\textbf{Optimization Problem.} The serving of each model $m$ in our scenario demands a configuration of batch size, device selection, GPU selection, and execution time $[bz_{m, g}, d, g, t]$. 
The SOTAs \cite{zeng2020distream, hu2021rim, nigade2022jellyfish} do not provide temporal scheduling ($t \in [-\infty, +\infty]$), leading to severe co-location interference and unpredictable performance \cite{Wu2022colocationinference}.
Instead, \systemname{} performs temporal scheduling to reduce interference and guarantee that models perform as expected.
Each configuration results in a batch inference latency of  $\mathcal{L}_{m|bz_{m, g}, d, g, t}$.
Since all batched queries finish at the same time \cite{crankshaw2020inferline}, the latency of each is averaged across the batch as $\mathcal{L}^{\text{infer}}_{m} = \frac{\mathcal{L}_{m|bz_{m,g}, d, g, t}}{bz_{m,g}}$.

\par
Furthermore, there is a latency in transferring the output of the previous model, $m_{prev}$, to $m$, which we define as $\mathcal{L}^{\text{io}}_{m} = \frac{\text{size}(In_m)}{BW}$, where $I_m$ is the input of $m$  and $BW$ is the bandwidth of the transfer medium between two devices. For two models located on the same device, the bandwidth only depends on the device's hardware, which can be considered a large constant $\epsilon$, and the latency can be negligible.
However, for two models located on two different devices, the bandwidth is subjected to the network conditions.
Thus, the average latency for each query of $m$ is $\mathcal{L}_{m}=\mathcal{L}^{\text{infer}}_{m} + \mathcal{L}^{\text{io}}_{m}$. Thus, we can have:
\begin{equation}
    \mathcal{L}_{m}  = \frac{\mathcal{L}_{m|bz_{m,g}, d, g, t} }{bz_{m,g}} + 
                                                \begin{cases}
                                                            \frac{\text{size}(In_m)}{\epsilon_{d_m}},& d_{m} = d_{m_{prev}}\\
                                                            \frac{\text{size}(In_m)}{BW_{d_m,d_{m_{prev}}}}, & \text{otherwise}
                                                \end{cases}
\end{equation}
The pipeline's average latency
is defined as $\mathcal{L}_p = \sum^{m \in p} \mathcal{L}_{m}$.
The \textbf{effective throughput}s of $p$ and the whole system are calculated as $\mathcal{G}_p = 1 / \mathcal{L}_p$ and $\mathcal{G} = \sum^{p \in \mathbf{P}} \mathcal{G}_p$, respectively. Our problem can be presented as a integer linear program (ILP):
\begin{equation}
    \label{eq:objective-func}
    \max_{bz_{m,g}, d, g, t} \sum^{p \in \mathbf{P}} \sum^{t \in T} \frac{1}{\sum^{m \in p}\mathcal{L}_{m}}
\end{equation}
which is subject to the following spatiotemporal constraints:

\textbf{a) The SLO of all requests is met even in this worst-case scenario} as the first request in each batch waits the longest for the batch to fill, experiencing the highest latency. 

\begin{table}[t]
\centering
\caption{Summary of frequently used notations.}
\scriptsize
\vspace{-0.5em}
    \label{Tab:notation}
\begin{tabular}{@{\hspace{1pt}}m{1.8cm}|@{\hspace{3pt}}m{6.5cm}@{}}
  \hline
  \textbf{Notation}& \textbf{Description} \\\hline
  $p, P$&  pipeline $p$ and list of all pipelines including $p$ \\\hline
  $m$, $m_{prev}$& model $m$ and the upstream model of $m$ \\\hline
  $d, D$& host device $d$ and list of all host devices including $d$ \\\hline
  $d_m$ & device that hosts an instance of model $m$ \\\hline
  $g, G_d$& GPU $g$ and list of available GPUs on device $d$\\ \hline
  $M_g$ & Total memory available on GPU $g$ \\\hline
  $bz_{m, g}$, $BZ_{m, g}$& optimal and list of available batch sizes of $m$ on $g$\\ \hline
  $In_m$, $Out_m$ & input and output of $m$ (e.g., image - bounding box)\\ \hline
 $\mathcal{L}_{m|bz_{m, g}, d, g, t}$& Inference latency of $m$ under $[bz_{m, g}, d, g, t]$\\ \hline
 $\mathcal{L}^{\text{io}}_{m}$,  $\mathcal{L}^{\text{infer}}_{m}$,  $\mathcal{L}_{m}$ &  Average IO, inference, and total latency of $m$\\\hline
 $W_\text{m}, I_\text{m}, W_\text{g}, I_\text{g}$ & weight/intermediate memory of model $m$ and all models on $g$ \\ \hline
 $U_{m, g}$, $U_g$& Utilization rates of $m$ on GPU $g$ and the current rate of $g$\\\hline
 
 \end{tabular}
 \vspace{0.5em}
\end{table}

\vspace{-0.2cm}
\begin{equation}
    \label{eq:slo-constraint}
    \sum_{m \in p}\mathcal{L}_m . bz_{m,g} = \sum_{m \in p}\mathcal{L}_m^{\text{worst}} \leq SLO_p, \forall p \in P    
\end{equation}

\textbf{b) The total memory consumption of all models on $g$, including persistent weights $W_m$ and temporary intermediate inference outputs $I_m$ \cite{Cox2021interlayers}, does not exceed $g$'s capacity}. This ensures no model crashes during runtime.
\begin{equation} 
    \label{eq:mem-constraint} \sum_{m \in g} W_m + I_m \leq M_g, \ \forall d \in D 
\end{equation}

\textbf{c) The current GPU utilization rate $U_g$ -- the sum of individual models' GPU utilization rates $U_{m_g}$ -- does not exceed GPU's maximum compute capability}. This ensures all models yield expected performances.
\begin{equation}
    \label{eq:capability-constraint}
    \sum_{m \in g} U_{m, g} = U_g \leq U_g^\text{max}, \forall g
\end{equation}
\par
Since solving the ILP problem is NP-hard, finding an optimal solution in real-time, especially under resource constraints at the Edge, is impractical.
Instead, we break it into two sub-problems: \textbf{cross-device workload distribution and co-location inference spatiotemporal scheduling}, whose solutions will be detailed in the next section.
For convenience, the notations are summarized in \autoref{Tab:notation}.


\section{\systemname's Design}
\label{sec:system-design}

\subsection{Overall Architecture}
\label{subsec:system-architecuture}

\par
\systemname{}'s architecture (\autoref{fig:overall-arch}) comprises 3 main components: a \textbf{Controller} that oversees system-wide scheduling and resource allocation; a \textbf{Device Agent} responsible for running containerized inference models on each device; and a \textbf{Knowledge Base} (KB) to store system-wide metrics. 
The operation cycle of \systemname{} contains 5 steps:

\circled{1} Each round starts with pipeline generation upon user requests.
The Controller collects network/workload statistics and model/device profiles necessary for scheduling from \textbf{KB}.

\circled{2} Then the \textbf{Controller} runs \cwd{} (\autoref{fig:workload-distributor}) to select batch sizes, host devices, and container instance counts for pipeline models to maximize throughput while meeting SLOs.

\circled{3} The \textbf{Controller} runs \coral{} to calculate timings and perform spatiotemporal scheduling for all scheduled models from step 2 to minimize the interference among co-located models.

\circled{4} Once finished, the scheduling results are communicated to the \textbf{Agent} on each device. It then deploys the models in containers, equipped with an inference engine (e.g., TensorRT) and managed by a container engine like Docker. The \textbf{Agent} also monitors the containers' metrics throughout the run time.

\circled{5} Container and device operation metrics including workloads, resource usage, and network bandwidth are monitored, and pushed to \textbf{KB} for monitoring and scheduling purposes.

\par
During run time, when containers reach throughput limits, \circled{2'} the \textbf{Auto Scaler} is triggered to scale up the containers by cloning to increase capacity.
These container \textit{instances} are then removed when demand drops.

\begin{figure}[t!]
    \begin{center}
    \includegraphics[width=\linewidth, trim=0cm 0cm 0cm 0.15cm, clip]{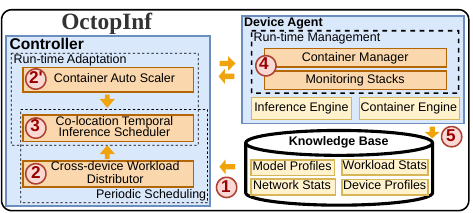} 
    \vspace{-1.5em}
    \caption{The overall architecture of \systemname{}}
    \label{fig:overall-arch}
    \end{center}
    \vspace{-0.5em}
\end{figure}

\subsection{Cross-Device Workload Distributor -- \cwd{}}
\label{subsec:workload-scheduling}

\par
Selecting the optimal batch sizes and workload distribution can eliminate bottlenecks within the pipeline in response to workload changes and network variations (\autoref{fig:workload-distributor}).
However, due to the large number of configurations, finding an optimal solution is prohibitively expensive.
To efficiently navigate the large search space, we propose a \textit{workload-aware} greedy algorithm based on the naive greedy min-max cost optimization algorithm proposed in~\cite{crankshaw2020inferline}, which treats all models in a pipeline similarly.
By incorporating observations and insights into workload characteristics such as burstiness and IO ratio, we can choose better starting points and quickly remove unfruitful configurations from consideration.

\textbf{\textit{Observation 1: Inference workloads are inherently bursty, with fluctuating intensity and varying from model to model depending on content dynamics}.}
For example, an object detector suddenly detects a crowd, generating many human boxes, which in turn creates a surge of traffic to a downstream face detector.
These surges can become even more pronounced during rush hour.
Also, this burstiness can propagate to different downstream models, affecting overall performance.\\
\textit{$\Rightarrow$ Insight 1: It is beneficial to prioritize larger batches for models with bursty workloads, which increases potential throughput. Additionally, since bursty workloads fill the batch quickly, each request has a shorter waiting time, which reduces the risk of SLO violations.}

\par
\textbf{\textit{ Observation 2: Higher network traffic between Edge devices and the server raises SLO-violation risks.}}
Optimized batch selections enable adaptation to workload variations, removing computation bottlenecks.
Still, unstable networks can be a bottleneck in the pipeline, slowing the whole pipeline.
\\
\textit{$\Rightarrow$ Insight 2: Placements should be chosen to minimize network traffic bottlenecks, improving system stability. 
For instance, during rush hour, an object detector capturing a high number of bounding boxes should not placed at the Edge being the split point because this generates higher network overhead than actually sending the raw frame to the server}.

\par
\textbf{\textit{Observation 3: More split points also increase the risk of SLO-violations.}}
Multi splits offer flexibility in model placement (e.g., \textit{edge$\rightarrow$server$\rightarrow$edge$\rightarrow$server}), but each additional split increases network traffic, raising the risk.\\
\textit{$\Rightarrow$ Insight 3: Number of pipeline splits should be minimized.}

\begin{figure}[!t]
    \begin{center}
    \includegraphics[width=\linewidth, trim=0cm 0cm 0cm 0.23cm, clip]{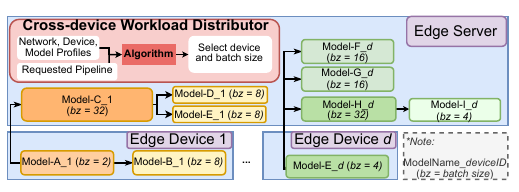} 
    \vspace{-0.6cm}
    \caption{\cwd{} searches for the near-optimal model deployments (location and batch size), considering requirements, network profiles, and device resources.}
    \label{fig:workload-distributor}
    \end{center}
    \vspace{-0.2cm}
\end{figure}

\begin{algorithm} [!t]
    \SetKwInput{Input}{Input}
    \SetAlFnt{\footnotesize} 
    \footnotesize 

    \SetKwFunction{Distribute}{ToEdge}
    \SetKwProg{Fn}{Function}{:}{}

    \SetKwFunction{cwdalgo}{cwd}
    \SetKwProg{Fn}{Function}{:}{}
    \Fn{\cwdalgo{}}{

        \For{Pipeline $p \in unscheduledPipelines$} {
            \For{Model $m \in p$} {
                $d_m = \text{server}$ ; $g$ = server\_gpu;
                 $bz_{m, g} = 1$ \\
                $\text{numInstances}_m = 1$; AddInstances($m$) \tcp*[h]{minimum cfg}
            }

            $p_\text{s} = $ Sort $m \in p$ \textbf{descendingly} on \textbf{burstiness} \tcp*[h]{Insight 1}\\
            \Repeat{$\text{tmp\_cfg}_p$ = NULL}{
                $\text{tmp\_cfg}_p$ = NULL\\
                \For{Model $m \in p_\text{s}$} {
                    $bz_{m, g } \ *= \ 2$; ReduceInstances($m$) \tcp*[h]{min=1}\\
                    \If{$\text{EstLat}(p) > \text{SLO}_p / 2$} {
                        $bz_{m, g} = bz_{m, g} / 2$; AddInstances($m$) \\
                        
                    } \Else{
                        \If{$\text{EstThrpt}(p) > \text{max}$} {
                            $\text{cfg}_m$ = $[d_m, g, bz_{m,g}, 0]$ \tcp*[h]{set new cfg}\\
                            Update max; Add $\text{cfg}_m \rightarrow \ \text{cfg}_p$
                            
                        } 
                        
                    }
                    
                }
                
            }
            \Distribute{$p[0]$} \\
            Add $p$ to $scheduledPipelines$
            
        }
        \textbf{return} $scheduledPipelines$\\
        
    }

    \SetKwFunction{Distribute}{ToEdge}
    \SetKwProg{Fn}{Function}{:}{}
    \Fn{\Distribute{$m$} \tcp*[h]{DFS-style traverse}}{
        $\text{cfg}_p$ = Find a new configuration only for $m$.\\
        \If{$\text{cfg}_p$ = NULL} {
            \textbf{return}
            
        }
        \For{$m_{next}$ \textit{sorted \textbf{ascendingly} on \textbf{burstiness}}} {
                \Distribute{$m_{next}$} \tcp*[h]{Insight 1}

        }
        \If{$\text{Overhead}(In_m) \cdot \alpha < \text{Overhead}(Out_m) \ \&\&$ \text{downstreams are \textit{not }on Edge}} { \tcp*[h]{Insights 2 and 3}\\
            Revert $\text{cfg}_m$
            
        }
        
    }   
    
    \caption{Cross-device Workload Distributor}
    \label{algo:workload-distributor}
\end{algorithm}

\par
The operation of \cwd{}'s algorithm (\autoref{algo:workload-distributor}) incorporating the above insights is detailed as follows. \cwd{} begins scheduling workloads for each pipeline $p$ by initializing minimal configurations for all models on the server and adding active instances to match incoming request rates (lines 3-5).
Then, the models are sorted in descending order based on their burstiness, measured by the coefficient of variation of inter-request arrival times (line 6). Next, \cwd{} greedily explores the configuration search space, increasing the batch size of each model $m$ starting with the burstiest (lines 9-10). 
This follows \textbf{\textit{Insight 1}} to maximize the throughput-enhancing capability of higher batch sizes.
Though, due to the nature of batched inference, each request experiences a higher latency, the burstiness helps minimize it by reducing the waiting time to fill the batch.
Next, thanks to the higher throughput, the number of instances of $m$ can be reduced to conserve resources (line 12).
However, if the new \textit{temporary} configuration violates the pipeline's SLO, it is dropped.
Otherwise, if the configuration improves throughput without violating the end-to-end SLO of the pipeline, it is adopted as the new configuration for $m$ (lines 14-16).
This exploration process continues until no better configuration is found for $p$ (line 17).

\par
Next, to distribute the workloads, we propose a DFS-style function, \Distribute{} which incrementally distributes the models to the edge device where the data source is located.
This follows \textbf{\textit{Insights 2 and 3}} to choose a minimal number of split points, each of which reduces the network traffic and overhead.
\Distribute{} begins with the first model (line 18) and applies a similar process as above to explore new configurations, though the search is constrained to avoid costly explorations (line 22).
If a suitable configuration is found, it continues traversing downstream, temporarily treating each model $m$ as a split point (lines 25-26).
In particular, $m$ is first assumed to fit \textit{Insight 2}, significantly reducing network traffic.
On the return, \Distribute{} tests this assumption by evaluating $m$'s input-output size ratio which represents trade-offs between receiving inputs from the network and transmitting the output.
Particularly, if $m$'s output does not exceed that of its input data multiplied by a factor $\alpha$, there are network benefits from placing $m$ at the Edge. However, if the condition does not hold (line 27, upper) and no downstream models serve as better split points (i.e., they are still on the server) (line 27, lower), \Distribute{} reverts $m$'s configuration and returns it to the server (line 28).
Additionally, \Distribute{} leverages \textit{Insight 1} by attempting to move less bursty models to the edge first, as their outputs are less likely to cause network bottlenecks.

\subsection{Co-location Inference Spatiotemporal Scheduler -- \coral{}}
\label{subsec:temporal-scheduling}

\begin{figure}[t!]
    \begin{center}
    \vspace{-0.1cm}
    \begin{subfigure}{\linewidth}
        \includegraphics[width=\linewidth, trim=0cm 0cm 0cm 0.1cm, clip]{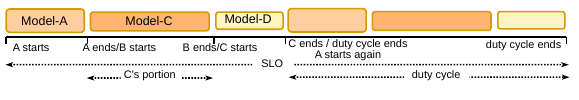} 
        \vspace{-0.9cm}
        \caption{Execution timings of a simple pipeline.}
        \label{subfig:pipeline-timing}
    \end{subfigure}
    \vspace{0.5cm} 
    \begin{subfigure}{\linewidth}
       \includegraphics[width=\linewidth, trim=0cm 0cm 0cm 0.1cm, clip]{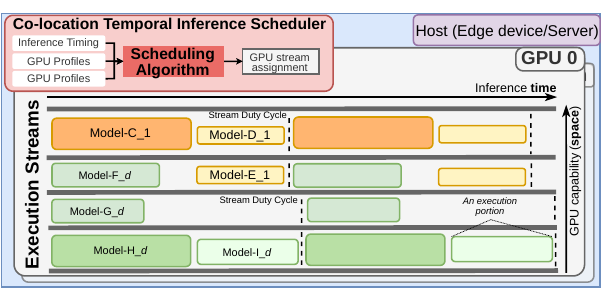} 
       \vspace{-0.7cm}
        \caption{Overview of co-location spatiotemporal scheduling.}
        \label{subfig:temporal-scheduler}
    \end{subfigure}
    \vspace{-1cm}
    \caption{Co-location Inference Spatiotemporal Scheduler (\coral)}
    \label{fig:temporal-scheduler}
    \end{center}
    \vspace{-0.1cm}
\end{figure}

\par
\par After completing \autoref{algo:workload-distributor}, \systemname{} retrieves the list of pipelines and their model instances, $scheduledPipelines$.
\systemname{} then invokes \coral{} to enable multi-model GPU sharing, aiming to maximize utilization while ensuring models operate without resource starvation.
To achieve this, \coral{} addresses challenges in navigating the dual dimensions of \textbf{time} and \textbf{space} (i.e., GPU hardware capacity at any moment).
We introduce the \textit{\textbf{inference stream}} (or simply \textbf{stream} abstraction to enable spatiotemporal GPU sharing and propose a best-fit scheduling algorithm leveraging streams.

\subsubsection{Inference Stream:}
\par
A GPU's inference capacity is divided into multiple concurrently executed streams, as shown in \autoref{fig:temporal-scheduler}.
Each stream represents a temporal sequence of model executions.
Each model execution occupies a \textit{portion}, whose length and width represent the execution time and required computation capability, respectively.

\par
When considering a pipeline, the end-to-end latency can be broken into the latencies of its models (\autoref{eq:objective-func}), each of which occupies a portion of streams.
The length of each portion can be calculated based on their selected batch sizes and batch inference profiles.
The execution sequence is arranged according to $p$'s DAG following a natural order ensuring downstream models immediately process data generated by their upstream.
For example, scheduling Model-D before C would waste D's portion, as its input has not yet been generated by C (\autoref{subfig:pipeline-timing}).
Each stream also has a \textbf{duty cycle}, which lasts for half of $p$’s SLO.
After D’s portion ends, GPU access is cycled back to A, which processes the new requests that arrive during the executions of C and D.

\begin{algorithm} [t!]
    \SetKwInput{Input}{Input}
    \SetKwInOut{IO}{Output}
    \SetAlFnt{\footnotesize}
    \footnotesize
    \DontPrintSemicolon
    
    \SetKwFunction{coralalgo}{CORAL}
    \SetKwProg{Fn}{Function}{:}{}
    
    \SetKwFunction{Main}{Main}
    \SetKwProg{Fn}{Function}{:}{}
    \Fn{\Main{}}{
        \footnotesize
        $instanceNumber = 0$ \\
        \While{Has unscheduled instances} {
            \For{$p \in scheduledPipelines$} {
                \For {$m \in p$} {
                    \If{$m$ has $instanceNumber$} {
                        \coralalgo{$m$, $instanceNumber$}
                        
                    }
                    
                }
                
            }
        
            $instanceNumber++$
            
        }
        
    }
    
    \Fn{\coralalgo{$m$, $instanceNumber$}}{
        \footnotesize
        $r = m[\textit{instanceNumber}]$ \\
        \For{portion $pt$ in $FreePortions$} {
            $s = pt_\text{stream}$;
            $g$ = $s_\text{gpu}$;\\
            $w_g = W_g + W_m$ \tcp*[h]{total weight memory}\\ 
            $i_g = I_g - I_\text{s} + \text{max}(I_\text{s}, I_m)$ \tcp*[h]{total intermediate memory}\\
            $u_g = \sum_{s' \in g} \max_{r' \in s} (U_r', U_r)$ \tcp*[h]{total GPU utilization}\\
            \If{$(pt_\text{start} \leq m_\text{start}  \ \&  \ pt_\text{end} \geq m_\text{end} )$ \& \tcp*[h]{fully available}\\
            \ \ \   $(w_g + i_g \leq  M_g$ \& $u_g \leq U_g^\text{max})$ \& \tcp*[h]{resource sufficiency} \\
            \ \ \   $(\text{duty\_cycle}_\text{r} \geq \text{duty\_cycle}_\text{s}$)}{
                \If{$\text{duty\_cycle}_\text{s}$ = 0} {
                    $\text{duty\_cycle}_{s} = \text{duty\_cycle}_\text{r}$ \tcp*[h]{update stream's}
                    
                }
                
                $\text{stream}_r$ = $s$;
                $\text{duty\_cycle}_{r}$ = $\text{duty\_cycle}_\text{s}$ \tcp*[h]{stats}\\
                $W_\text{g} = w_\text{g}$; $I_\text{g} = i_\text{g}$; $U_g=u_g$ \tcp*[h]{update resource usage}\\
                $\text{selected\_pt, freed\_pts} = \text{DividePortion}(pt)$\\
                UpdateList(freed\_pts, $FreePortions$)\\
                \return{selected\_pt} \\
                
            }
            
        }
        
        \return{\textit{not found}}
        
    }
    
\caption{Co-location Inference Spatiotemporal Scheduler}
\label{algo:temporal-scheduler}
\end{algorithm}

\subsubsection{Spatiotemporal Scheduling Algorithm:}
\par
The operation of \coral{}'s scheduling algorithm is shown in \autoref{algo:temporal-scheduler}.
After \cwd{} finishes, \coral{} retrieves the list of pipelines with scheduled workloads, $scheduledPipelines$.
It loops through these and calls \coralalgo{} for unscheduled instances (lines 3-7).
In each iteration, only one instance per model is scheduled to ensure fairness, so all pipelines have at least one active instance when priorities are equal.
Complex scenarios with different priority policies can easily be integrated into \coral{}.

\par
\coral{}'s objective is to schedule as many instances on available GPU streams as possible, efficiently stacking execution portions one after another to minimize gaps, which waste resources.
For each container instance $r$ of model $m$, \coral{} searches for the best-fitting portion $pt$ among the list of free portions (line 11).
For each, \coral{} estimates the potential resource consumptions of GPU $g$ if $r$ is scheduled on $pt$ (lines 12-15).
Then \coral{} evaluates $pt$ on three conditions:

\textbf{(1) $pt$ fully contains $r$'s portion with minimal empty space}, which ensures stream $s$ is fully available during the execution of $r$ (line 16).

\textbf{(2) GPU $g$ has sufficient memory and compute capacity for $r$ (line 17)} to satisfy \autoref{eq:mem-constraint} and \ref{eq:capability-constraint}. This ensures no model crashes due to insufficient memory and there is enough computation to avoid co-location interference.

\textbf{(3) Pipeline $p$'s duty cycle must be longer than that of stream $s$ (line 18)}. This ensures that having $r$ on does not prolong other models' duty cycles and causes them to violate their SLOs, satisfying \autoref{eq:slo-constraint}.

\par Once the best-fit portion is found, $r$ is assigned to stream $s$, the duty cycle of $s$ is adjusted, and $g$'s operational stats are updated (lines 19-22).
If $r$ cannot fully use the portion, the original $pt$ portion is divided and the leftover is added back to the $FreePortions$ list for other instances (lines 23-24), ensuring optimal GPU resource utilization. 

\subsection{Run-time Horizontal AutoScaler}
\par
During runtime, workload fluctuations in models can occur due to content dynamics.
While \systemname{} can generate near-optimal schedules, frequently running \cwd{} incurs higher scheduling overhead, especially given the large search space.
To address this, \systemname{} complement periodical scheduling with a quick response strategy for surges and dips.
When \systemname{} detects significant workload surges for model $m$, it triggers the \textit{AutoScaler} (\autoref{fig:overall-arch} create additional instances of $m$ and schedules them temporally as described earlier.
Conversely, unnecessary instances are removed, and their assigned portions are reclaimed.


\section{Evaluation}
\label{sec:evaluation}

\subsection{Experiment Setups}
\label{sub:experiment-setups}

\subsubsection{Testbed \& System Implementation:}
\label{subsubsec:system-implementation}

\par
We evaluate \systemname{} on a real-world testbed comprising a server with 4 NVIDIA RTX 3090 GPUs and 9 heterogeneous Jetson Edge devices (1 AGX, 5 Xavier NXs, and 3 Orin Nanos).
The Edge server and devices run Ubuntu 20.04 with CUDA 11.4.

\par
\systemname{} is implemented in over 25K lines of C++ at \href{\sourcecodelink}{\sourcecodelink}.
It adopts a container-based architecture, where each model runs within a container for seamless scaling and adaptation in dynamic environments.
Containers communicate via gRPC for efficient inter-service communication.
The \textbf{Controller} are run at the server. On each device that handles workloads (including the server), a \textbf{Device Agent} is run as a separate process to manage and monitor containers via the Docker API and NVIDIA driver APIs, reporting statistics to the PostgreSQL \textbf{Knowledge Base} (KB).
During scheduling, the \textbf{Controller} queries the \textbf{KB}, performs scheduling and allocation, and communicates decisions to the \textbf{Agents}, which enforces them on their containers.

\subsubsection{Experiment Pipelines:}
\label{subsubsec:evaluation-pipelines}
Our experiments are conducted using 2 pipelines depicted in \autoref{fig:pipelines}.
The pipeline end-to-end SLOs are set at 200ms for \textit{traffic} and 300ms for \textit{surveillance}.
These values remain the same for all experiments except in \autoref{subsubsec:slo-variation}, where we restrict the SLOs even further.

\subsubsection{Data:}
\label{subsubsec:evaluation-data}

\par
We collected nine 13-hour videos from real-world public online streams (e.g., \href{\earthcam}{\earthcam}) at 15 fps and 1280x720 resolution.
These streams, including 6 \textit{traffic} and 3 building \textit{surveillance}, contain human and vehicle targets and were chosen to represent a wide range of target distributions and content dynamics.
For the main experiments, we extracted 30-min segments from three different times of the day to capture varying content dynamics.
For full replicability, the code for data collection is also documented in the source code.
\subsubsection{Baselines:}
\label{subsubsec:baselines}
For fair comparisons, we implement three state-of-the-art Edge inference, namely \textbf{Distream} \cite{zeng2020distream}, \textbf{Jellyfish} \cite{nigade2022jellyfish}, and \textbf{Rim} \cite{hu2021rim}, on top of the same source code and also make slight adjustments to show their best performances:
\begin{itemize}[wide]
    \item All baselines do not provide any GPU scheduling. Thus, we implement a best-fit algorithm to spread models evenly based on resource consumption across GPUs.
    \item We adjust the static batch sizes for Distream and Rim to 4 at the edge, 8 at the server, and 2 for Object Det. This configuration shows the best performance.
    \item We implement lazy dropping of late requests to Distream and Rim to give them a higher effective throughput.
    \item Jellyfish does not consider pipelines but places multiple YOLOv5 versions at the Server. We match the number of downstream model instances to that of YOLOv5 versions with a static batch of 8 similar to Distream and Rim.
\end{itemize}

\subsubsection{Experiment settings:}
\label{subsubsec:experiment-settings}
\par
Each collected video is assigned to an edge device, streamed like a real data source (e.g., cctv camera).
To emulate the real-world network conditions, we extract network traces from an Irish 5G data set for all Edge devices~\cite{raca2020beyond}.
The time between scheduling periods is set to 6 minutes for all algorithms.
For fairness, the averaged results of 3 runs are reported for all experiments.

\subsection{Evaluation Metrics:}
The following three key metrics are used to evaluate our system and baselines:
\begin{itemize}[wide]
    \item \textbf{Effective throughput:} The main metric for \systemname{} is end-to-end effective throughput -- \textbf{\textit{the number of objects arriving on time at the sink every second}}. We compare effective throughput to total throughput, considering the difference as wasted computations as requests arrive later than the specified service-level objectives (SLOs).
    This wasted computation causes further delays and consumes additional energy.
    \item \textbf{End-to-end latency}: A slim latency distribution indicates reliable performance and ability to satisfy stringent demands.
    \item \textbf{Total memory allocation:} A smaller memory footprint indicates better resource utilization.
\end{itemize}

\begin{figure}[t!]
    \begin{center}
    \includegraphics[width=\linewidth,trim=0 0 0 0, clip]{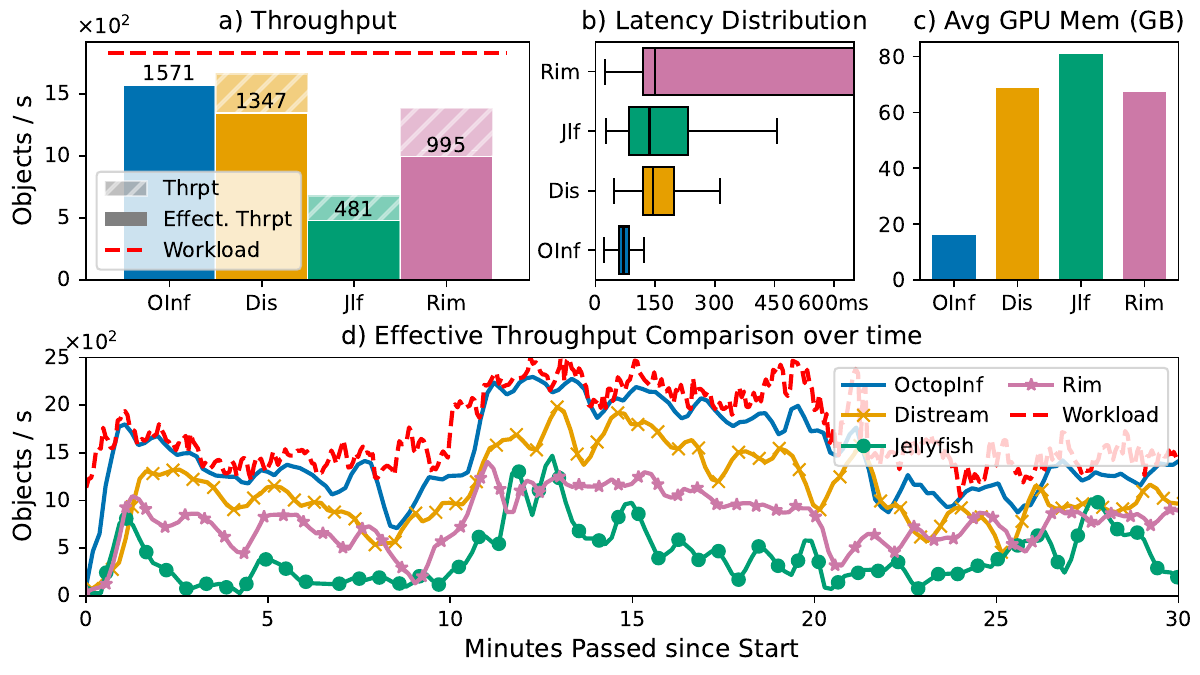} 
    \vspace{-1.5em}
    \caption{Overall performance comparisons under environmental dynamics}
    \label{fig:overall-performance}
    \end{center}
    \vspace{-0.5em}
\end{figure}

\subsection{Evaluation Results}
\label{subsec:evaluation-results}

\subsubsection{\textbf{Can \systemname{} outperform the baselines on real-world dynamic workloads and network conditions?}}
\label{subsubsec:overall-perforamnce}

\par
\autoref{fig:overall-performance}a and b show that \systemname{} significantly outperforms the baselines with higher effective throughput and shorter latency.
While Distream and Rim achieve throughputs close to \systemname{}, around 20\% and 30\% of their requests, respectively, violate their SLOs, rendering them wasted computation.
In terms of effective throughput, \systemname{} outperforms them by 16\% and 57\% respectively.
Also, \autoref{fig:overall-performance}d shows that \systemname{} finely adapts and matches the constantly changing workload throughout the entirety of the experiment.
These can be achieved thanks to \systemname{}'s effective scheduling.
First, \systemname{}, with fine-grained batch size selection, can distribute the workload more efficiently and avoid communication and computation bottlenecks, which slow down the processing of the whole pipeline and lead to SLO violation.
Second, with GPU spatiotemporal scheduling to allow orderly GPU access avoiding co-location interference, \systemname{}'s requests are processed on time according to calculation instead of being delayed due to resource contention. 
Jellyfish shows the worst throughput and 2nd worst latency due to its centralized architecture.
Particularly, Jellyfish relies resolution reduction to lower network latency. Yet these resized frames still constantly have to be transferred to the server requiring a significant bandwidth, which is unstable in real-world conditions.
Moreover, Rim shows the worst latency, because it attempts to move as many models to the Edge as possible without accounting for the workload dynamics and network conditions. 
More models on edge devices amplify the effects of co-location interference due to reduced parallel capabilities.

\begin{figure}[t!]
    \begin{center}
    \includegraphics[width=\linewidth,trim=0 0 0 0, clip]{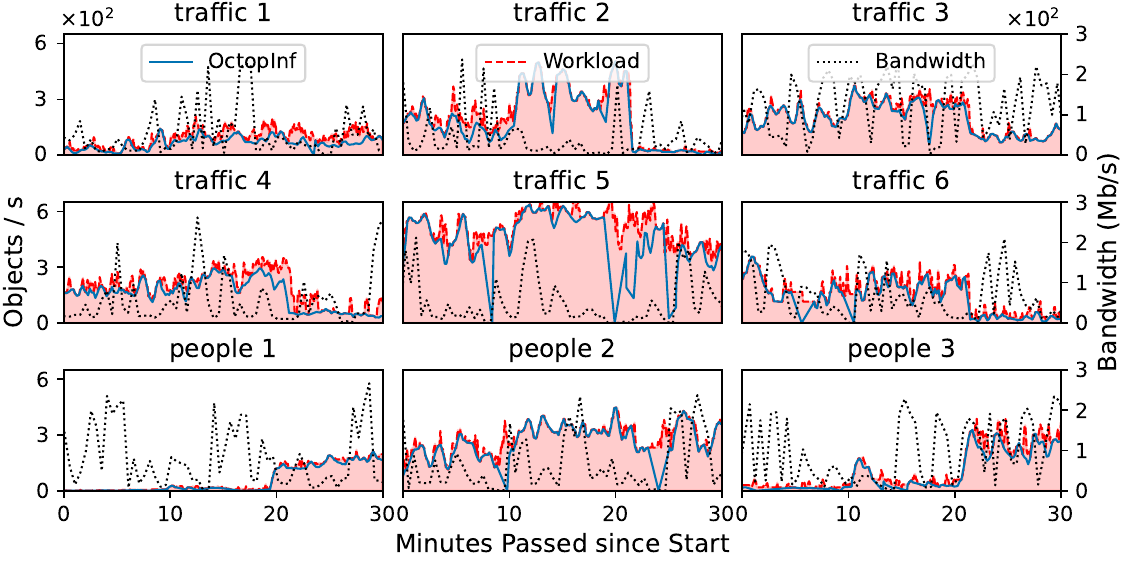} 
    \caption{Detailed workload, bandwidth and throughput patterns for \systemname{}.}
    \label{fig:individual-streams}
    \end{center}
    \vspace{-0.5em}
\end{figure}

\par As illustrated in \autoref{fig:overall-performance}c, \systemname{} utilizes significantly less memory compared to the baselines, primarily thanks to temporal sharing of the GPU.
When a model is idle, the GPU only needs to allocate memory for its relatively small weight.
In contrast, running a model demands substantial memory for I/O buffers and intermediate layers \cite{Cox2021interlayers}.
This issue is exacerbated when many models are deployed simultaneously on the server (e.g., 30 for Jellyfish) without proper scheduling.
Notably, Rim demonstrates the second lowest GPU memory usage as it seeks to accommodate as many models as possible at the Edge.

\begin{figure}[t!]
    \begin{center}
    \includegraphics[width=\linewidth,trim=0 10 0 10, clip]{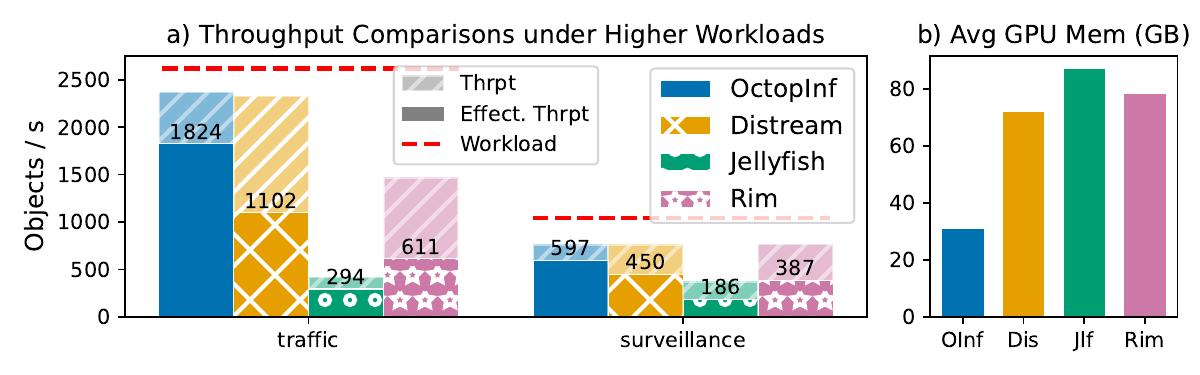} 
    \caption{Performance comparisons under higher workloads.}
    \label{fig:30-fps-experiment}
    \end{center}
    \vspace{-0.5em}
\end{figure}

\subsubsection{\textbf{How effectively can \systemname{} handle real-world workload dynamics and network conditions}?}
\label{subsubsec:network-bandwidth}

We assess \systemname{} using individual data sources under LTE bandwidth traces \cite{raca2020beyond}.
\autoref{fig:individual-streams} shows that \systemname{} effectively adapts to fluctuating workloads, maintaining throughput aligned with bandwidth changes.
Yet, there are instances of network disconnection (e.g., \textit{traffic 5} at the 19th and 25th minutes) throughput unavoidably falls to 0.
Otherwise, during low bandwidth periods, \systemname{} remains resilient, closely matching workload demands (e.g., \textit{people 2} at the 20th minute and \textit{people 3} at the 28th minute).

\par
This performance is achieved through the intelligent design of \cwd{} when considering the network instability.
From an analytics perspective, users prioritize identifying objects and their attributes, making throughput (\textit{objects/s}) a key metric.
However, content dynamics, such as changes in object shapes and sizes, can lead to variations in network traffic for the same throughput. To address this, \systemname{} considers the input-output size ratio, which assesses the trade-off between receiving inputs over the network versus transmitting outputs.
If yes, placing the model at the edge will certainly reduce network traffic from the edge to the server.
By incorporating this practical view, \systemname{} mitigates network bottlenecks, ultimately improving overall throughput.

\subsubsection{\textbf{Can \systemname{} scale to higher system-wide workloads?}}
\label{subsubsec:high-workload}
\par
We introduce an additional data source to each device to simulate scenarios where multiple cameras (e.g., different angles) are connected to the same device, effectively doubling the frame rate and system-wide workload to evalutate \systemname{}'s resilience.
\autoref{fig:30-fps-experiment} shows the baselines have significantly lower throughput ratios compared to \autoref{fig:overall-performance}, with even Distream only achieving a 50\% ratio.
This is because while the workload doubles, the relative burstiness ratio among more actually quadruples.
It causes Jellyfish to encounter severe network bottlenecks, being able to complete only 14\% of the requests.
Distream and Rim, which do not have dynamic batching, still fare better with workload distribution.
Yet, due to fixed batch sizes, their coarse-grained distribution overloads edge devices leading to poor performances and does not account for network traffic, though less severely compared to Jellyfish.
This proves the validity of our design, which uses workload dynamics (e.g., burstiness) to navigate dynamic batching for workload allocation.
Additionally, higher workload means more hardware usage (\autoref{fig:30-fps-experiment}b), leading to more severe co-location interference.
This further showcases the positive effects of \coral{}'s spatiotemporal scheduling.

\subsubsection{\textbf{Can \systemname{} flexibly adapt to stricter SLO demands?}}
\label{subsubsec:slo-variation}

\begin{figure}[t!]
    \begin{center}
    \includegraphics[width=\linewidth,trim=0 11 0 9, clip]{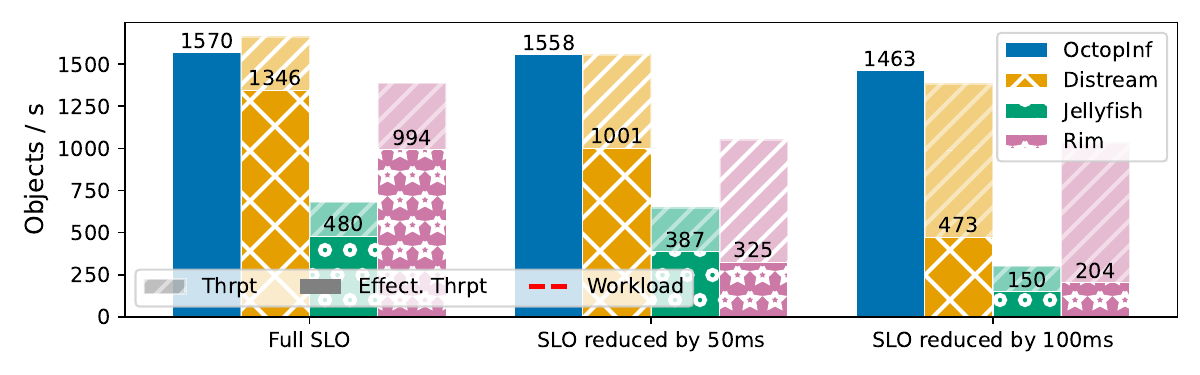} 
    \caption{Throughput comparisons under stricter SLO demands.}
    \label{fig:slo-experiment}
    \end{center}
    \vspace{-0.5em}
\end{figure}

\begin{figure}[t!]
    \begin{center}
    \includegraphics[width=\linewidth,trim=0 14 0 10, clip]{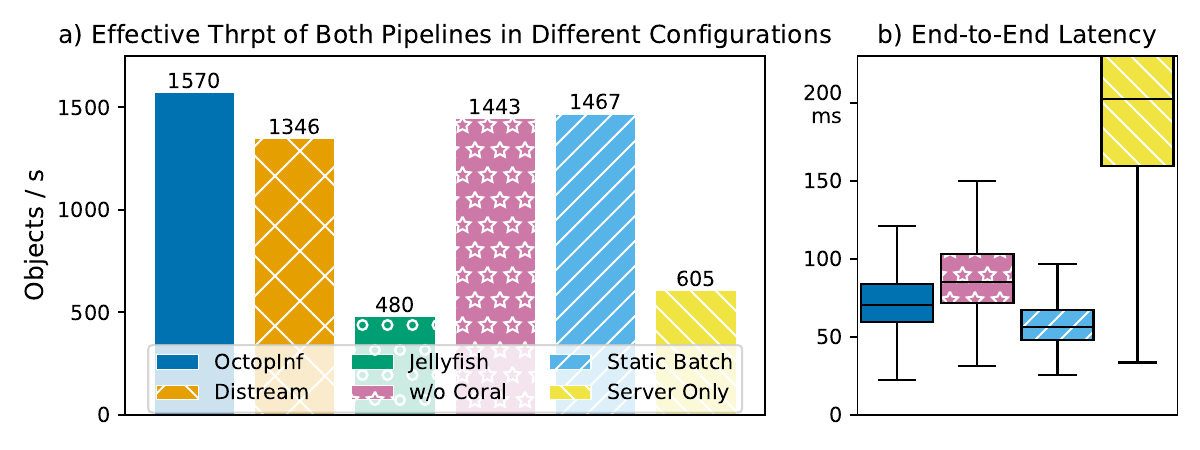} 
    \caption{Ablation Study of \systemname{} in comparison to Jellyfish and Distream.}
    \label{fig:ablation-study}
    \end{center}
    \vspace{-0.5em}
\end{figure}
\par
In this experiment, we gradually reduce the pipeline SLOs by 50-100ms from their original value of 200ms and 300ms (\textit{traffic} and \textit{surveillance} pipelines respectively) to test how well the systems adapt to stricter demands.
\autoref{fig:slo-experiment} demonstrates that \systemname{} maintains its performance and continues to outperform the baselines, which experience significant performance degradation.
Compared to full SLOs, the baselines' \textit{effective throughputs} drop  3-5$\times$ (Rim) because despite their attempts to adjust scheduling and allocation to the new requirements, \textit{co-location interference} has a greater impact at tight latency targets.
Moreover, beside Jellyfish which still suffers heavily from network bottlenecks, Distream and Rim have less options to reduce the latency due to their fixed batches (similar to assembling clunky latency chunks).
On the other hand, \systemname{} can effectively adjust its batch sizes and balance between latency and throughput (similar to having access to a much wider variety of chunks).
With SLOs reduced by 100ms, \systemname{} achieves staggering 7 and 10$\times$ throughput improvements compared to Rim and Jellyfish.

\subsubsection{\textbf{How does each component contribute to \systemname{}'s overall performance?}}
\label{subsubsec:ablation-study}
To quantify the impacts of different components, we conduct an ablation study by \textit{turning off} specific features individually, leading to 3 settings: \textit{w/o Coral} (no spatiotemporal scheduling), \textit{Static Batch} (fixed batches with spatiotemporal scheduling), and \textit{Server Only} (dynamic batching for server-only deployments with spatiotemporal scheduling).

\par
As shown in \autoref{fig:ablation-study}, the best performance is achieved when all components of \systemname{} are active.
The performance decreases about 10\% for \textit{w/o Coral}.
Here, many models simultaneously submit their inference workloads to CUDA, which treats each workload as a set of computation kernels instead of a single unit that requires to be completed on time.
To ensure fairness, CUDA alternatively schedules hardware for kernels of different models, leading to higher latency for all models (\autoref{fig:ablation-study}b).
\systemname{} schedules and reserves an exclusive \textit{portion} for each model to ensure that it has enough resource to complete on time.
A smaller performance drop occurs with \textit{Static Batch}, \systemname{} still maintains a high throughput by dynamically distributing the workloads following \cwd{}'s \textit{Insights 2 and 3}.
In this case, the present of spatiotemporal scheduling also helps mitigate the issue of co-location interference.
Interestingly, this setting achieves less throughput but has lower latency \autoref{fig:ablation-study}.
This proves that the latency-throughput balancing ability from dynamic batching is essential.
It is also worth noting that both \textit{w/o Coral} and \textit{Static Batch} still outperform the baselines.
The enormous performance drop occurs with \textit{Server Only}.
Similar to Jellyfish, it suffers from severe network bottlenecks and thus high latency, though thanks to spatiotemporal scheduling it still outperforms Jellyfish.
This shows the effectiveness and necessity of \systemname{}'s contributions to solving the challenges of EVA inference serving.

\subsubsection{\textbf{Is \systemname{} reliable for long-term operations?}}
\label{subsubsec:longterm-reliability}

\par
In this experiment, we test the long-term reliability of \systemname{} using the complete dataset, which includes nine 13-hour video segments. Each video records continuously from 9 AM to 10 PM, capturing the full range of content dynamics at the Edge.

\par
\systemname{} maintains high \textit{effective throughput} for both pipelines throughout the entire runtime, following patterns that align with human circadian rhythms as depicted in \autoref{fig:long-term-experiment}.
The \textit{traffic} pipeline starts with lower throughput, peaks around the 450th minute (3:30 PM), and tapers off by the 600th minute (8 PM).
During this period, the results show minimal throughput reduction compared to the workload. 
This demonstrates that \systemname{} is stable, adaptable to real-world environments over extended periods, and capable of maintaining optimal performance with minimal resource usage.

\begin{figure}[t!]
    \begin{center}
    \includegraphics[width=\linewidth,trim=0 10 0 10, clip]{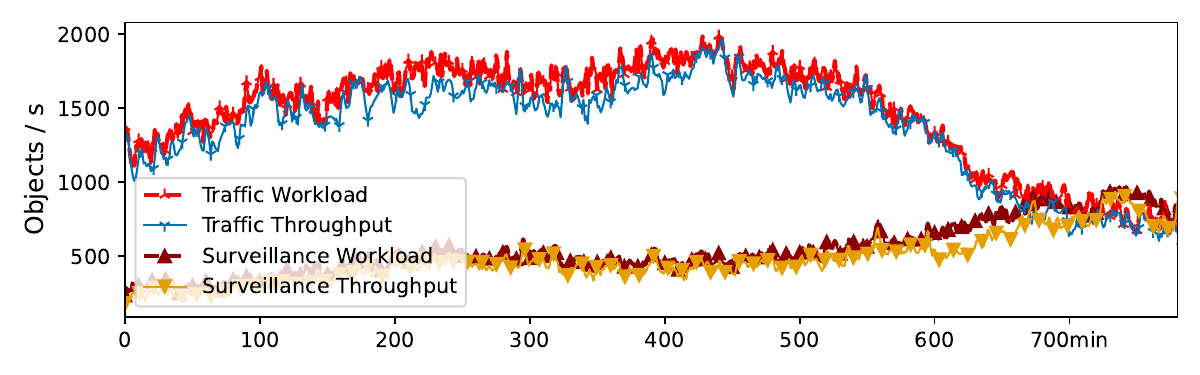} 
    \caption{Effective throughput of \textit{traffic} and \textit{surveillance} plotted over 13h.}
    \label{fig:long-term-experiment}
    \end{center}
    \vspace{-0.5em}
\end{figure}

\section{Discussion}
\label{sec:discussion}

\subsubsection{\textbf{Scalability}:} We discuss how \systemname{} can work at scale by reducing the complexity and being able to seamlessly integrate major inference platforms and model architectures.
\label{subsubsec:scalability}
\hfill

\textit{Complexity Reduction.}
Let $D$, $G$, $M$, and $BZ$ represent the numbers of devices, GPUs, models, and available batch sizes for each model, respectively.
The inference serving problem's optimization space has a complexity of $\mathcal{O}(D \cdot (BZ \cdot G)^M)$, rendering the optimal solution computationally infeasible.
\systemname{} simplifies this by decomposing it into two \textbf{\textit{manageable}} subproblems: \textit{cross-device workload distribution} and \textit{co-location inference spatiotemporal scheduling}.
For workload distribution, a \textit{workload-aware} greedy algorithm approximates the solution with $\mathcal{O}(D \cdot M \cdot BZ)$.
For spatiotemporal scheduling, the \textit{stream} abstraction reduces complexity to $\mathcal{O}(M \cdot PT)$, where $PT$, capped by the number of model instances, denotes the maximum number of free portions.
This approach yields an overall complexity of $\mathcal{O}(D \cdot M \cdot BZ + M \cdot PT)$.
While only near-optimal, \systemname{} surpasses SOTA methods and, critically, operates in real-time -- vital for EVA systems.

\par
\textit{Inference Platform Heterogeneity.}
Scalability at the Edge requires generalizability across diverse hardware and software platforms. \systemname{} achieves this through two key concepts: \textit{batched inference} and \textit{inference stream}.
The benefits of \textit{batched inference} are supported by major platforms, including GPUs with TensorRT~\cite{Zhou2022trt}, CPUs with OpenVINO~\cite{Demidovskij2020openvino}, and TPUs with TensorFlow~\cite{Kochura2020tpu}.
\systemname{}'s \textit{inference stream} abstraction leverages low-level APIs for asynchronous and sequential execution, such as TensorRT streams, OpenVINO streams, and ONNX threads.
This design ensures \systemname{} can scale efficiently and adapt to dynamic Edge environments.

\par
\textit{Model Heterogeneity.}
While designed for EVA, \systemname{} can extend to other domains like time-series analysis with RNNs.
Batched inference improves RNN throughput, as shown in~\cite{Silfa2022rnn}, while their recurrent operations allow precise inference time estimation for scheduling.
These features enable \systemname{} to scale its workload distribution and spatiotemporal scheduling to diverse applications.

\subsubsection{\textbf{Co-location Interference among Streams}:}
\systemname{} mitigates co-location interference by ensuring the total resource consumption of models is within GPU capability enabling models' peak performances. However, these peaks are typically brief in terms of memory consumption and utilization, and not all models within the same streams require the same amount of memory and achieve similar utilization due to factors like model architecture and implementations.
Aligning these peaks more precisely could allow for more streams and better GPU utilization, but this requires a detailed analysis of workload patterns, which are highly diverse.

\subsubsection{\textbf{Future Work}:} We highlight the following future work to improve \systemname{}.
\textbf{First,} we plan to expand the testbed and integrate other inference platforms. Our first hardware target is Raspberry Pi devices using software frameworks ONNX and Tensorflow for their high compatibility.
\textbf{Second,} we aim to leverage fine-grained performance data as time series for a prediction model that estimates co-located model performance.
To accommodate the heterogeneity of model architectures, we plan to leverage techniques like Meta Learning and Test-time Adaptation (TTA) for quick adaptation to new architectures.
\textbf{Third,} we plan to enhance \systemname{} with fine-grained, second-scale local adaptation. Each model container will use Reinforcement Learning to learn workload patterns and make runtime decisions, such as adjusting batch sizes or request priorities, improving responsiveness to workload spikes.


\section{Related Work}
\label{sec:related-work}

\subsubsection{\textbf{Edge Video Analytics}} can be categorized into:
\hfill

\textit{Centralized Architecture: } dictates that video data is collected from multiple sources and processed at a central location.
To reduce overhead, approaches like VideoStorm \cite{zhang2017videostorm, Jiang2018chameleon, Tan2020FastVA, Zhang2018awstream, Li2020reducto} exploit the resource-quality tradeoff by adjusting parameters such as encoding quality, frame rate, and resolution.
Another strategy \cite{Kang2017noscope,Zhang2015vigil} uses models of varying accuracy, reserving high-precision models for complex tasks while employing cheaper ones for simpler tasks.
Jellyfish \cite{nigade2022jellyfish} combines both methods and provides a dynamic programming algorithm to handle network variability.

\textit{Distributed Architecture: } supports both device-server and device-device workload distributions.
While better suited to dynamic environments, it creates a larger search space, including the challenge of finding optimal \textit{split points} for workload partitioning.
DRLTO \cite{wang2022DRLTO} uses a fixed split and device-based analysis to optimize offloaded frames.
Rim \cite{hu2021rim} maximizes workload at one device to increase resource utilization while approaches like Distream \cite{Hung2018VideoEdge, liang2024splitstream, zeng2020distream} use stochastic methods to explore the search space.
EdgeVision \cite{gao2024edgevision} applies reinforcement learning to learn optimal configurations.

A common drawback of these systems is not fully utilizing an effective tool, \textit{dynamic batching}, to flexibly handle dynamic environments, as it expands the complex search space. Additionally, they fail to address the challenges posed by GPU execution of models, such as \textit{co-location interference}.

\subsubsection{\textbf{Inference Serving beyond the Edge}} serves a inference requests of various tasks at the Cloud \cite{shen2019nexus, Tan2021migserving, Choi2022Gpulet, Gujarati2020Clockwork, Zhang2023SHEPHERD, crankshaw2020inferline}.
These systems consider large GPU clusters where one can take several GPUs or a whole cluster to serve a high concentration of requests.
Although they do not address challenges at the Edge, such as resource constraints and dynamic environments (e.g. network), techniques such as \textit{dynamic batching} and GPU scheduling have inspired \systemname{}.

\section{Conclusion}
\label{sec:conclusion}

In this paper, we presented \systemname{}, a workload-aware inference serving system designed for Edge Video Analytics.
It uses dynamic inference batching combined with adaptive cross-device workload distribution to manage fluctuating workloads and introduces a novel co-location inference spatiotemporal scheduling algorithm to mitigate resource contention.
Our experiments demonstrate up to 10$\times$ improvement in \textit{effective throughput} compared to state-of-the-art baselines across various experiments, with a stable end-to-end latency distribution even under challenging conditions.
While our evaluation focused on Video Analytics, the proposed system can be extended to optimize any machine learning workload, with testing and optimization in those contexts left as future work.
With its low-level optimizations, \systemname{} is poised to be a key component of future Edge AI applications.


\bibliographystyle{IEEEtran}
\bibliography{paper}

\begin{thebibliography}{10}
\providecommand{\url}[1]{#1}
\csname url@samestyle\endcsname
\providecommand{\newblock}{\relax}
\providecommand{\bibinfo}[2]{#2}
\providecommand{\BIBentrySTDinterwordspacing}{\spaceskip=0pt\relax}
\providecommand{\BIBentryALTinterwordstretchfactor}{4}
\providecommand{\BIBentryALTinterwordspacing}{\spaceskip=\fontdimen2\font plus
\BIBentryALTinterwordstretchfactor\fontdimen3\font minus
  \fontdimen4\font\relax}
\providecommand{\BIBforeignlanguage}[2]{{%
\expandafter\ifx\csname l@#1\endcsname\relax
\typeout{** WARNING: IEEEtran.bst: No hyphenation pattern has been}%
\typeout{** loaded for the language `#1'. Using the pattern for}%
\typeout{** the default language instead.}%
\else
\language=\csname l@#1\endcsname
\fi
#2}}
\providecommand{\BIBdecl}{\relax}
\BIBdecl

\bibitem{bahl2020percom}
P.~V. Bahl, R.~Caceres, N.~Davies, and R.~Want, ``{Pervasive Computing at the
  Edge},'' \emph{IEEE Pervasive Computing}, vol.~19, no.~4, pp. 8--9, 2020.

\bibitem{Pasandi2020convince}
H.~B. Pasandi and T.~Nadeem, ``{CONVINCE: Collaborative Cross-Camera Video
  Analytics at the Edge},'' in \emph{2020 IEEE Int. Conf. on Pervasive
  Computing and Communications Workshops, PerCom Workshops 2020}.\hskip 1em
  plus 0.5em minus 0.4em\relax IEEE, 2020, pp. 1--5.

\bibitem{Hayashi2022traffic}
K.~Hayashi, A.~Hiromori, H.~Yamaguchi, M.~Suzuki, and T.~Kitahara,
  ``{Synthesizing Town-scale Vehicle Mobility from Traffic Surveillance
  Cameras: A Case Study},'' in \emph{2022 IEEE Int. Conf. on Pervasive
  Computing and Communications Workshops}.\hskip 1em plus 0.5em minus
  0.4em\relax IEEE, 2022, pp. 593--598.

\bibitem{Nguyen2023preacto}
T.-T. Nguyen, S.~Y. Jang, B.~Kostadinov, and D.~Lee, ``{PreActo: Efficient
  Cross-Camera Object Tracking System in Video Analytics Edge Computing},'' in
  \emph{2023 IEEE Int. Conf. on Pervasive Computing and Communications}.\hskip
  1em plus 0.5em minus 0.4em\relax IEEE, 2023, pp. 101--110.

\bibitem{Anjum2024surveillance}
K.~Anjum, T.~Chowdhury, S.~Mandava, B.~Piccoli, and D.~Pompili, ``{Leveraging
  On-Board UAV Motion Estimation for Lightweight Macroscopic Crowd
  Identification},'' in \emph{2024 IEEE Int. Conf. on Pervasive Computing and
  Communications (PerCom)}, 2024, pp. 11--17.

\bibitem{Sahu2023healthcare}
S.~K. Sahu, A.~L. Ruscelli, G.~Cecchetti, M.~Gharbaoui, and P.~Castoldi, ``{A
  perspective of telemedicine videostreaming systems for emergency care},'' in
  \emph{2023 IEEE Int. Conf. on Pervasive Computing and Communications
  Workshops and other Affiliated Events (PerCom Workshops)}, 2023, pp.
  122--127.

\bibitem{Kumrai2020activityrecognition}
T.~Kumrai, J.~Korpela, T.~Maekawa, Y.~Yu, and R.~Kanai, ``{Human Activity
  Recognition with Deep Reinforcement Learning using the Camera of a Mobile
  Robot},'' in \emph{2020 IEEE Int. Conf. on Pervasive Computing and
  Communications (PerCom)}.\hskip 1em plus 0.5em minus 0.4em\relax IEEE, 2020,
  pp. 1--10.

\bibitem{Bicocchi2012activityrecognition}
N.~Bicocchi, M.~Lasagni, and F.~Zambonelli, ``{Bridging vision and commonsense
  for multimodal situation recognition in pervasive systems},'' in \emph{2012
  IEEE Int. Conf. on Pervasive Computing and Communications}.\hskip 1em plus
  0.5em minus 0.4em\relax IEEE, 2012, pp. 48--56.

\bibitem{Jang2021pipeline}
S.~Y. Jang, B.~Kostadinov, and D.~Lee, ``{Microservice-based Edge Device
  Architecture for Video Analytics},'' in \emph{6th ACM/IEEE Symp. on Edge
  Computing, SEC 2021}.\hskip 1em plus 0.5em minus 0.4em\relax ACM, 2021, pp.
  165--177.

\bibitem{shen2019nexus}
H.~Shen, L.~Chen, Y.~Jin, L.~Zhao, B.~Kong, M.~Philipose, A.~Krishnamurthy, and
  R.~Sundaram, ``Nexus: A gpu cluster engine for accelerating dnn-based video
  analysis,'' in \emph{Proc. of the 27th ACM Symp. on Operating Systems
  Principles}, 2019, pp. 322--337.

\bibitem{Choi2022Gpulet}
S.~Choi, S.~Lee, Y.~Kim, J.~Park, Y.~Kwon, and J.~Huh, ``Serving heterogeneous
  machine learning models on {Multi-GPU} servers with {Spatio-Temporal}
  sharing,'' in \emph{2022 USENIX ATC}, 2022, pp. 199--216.

\bibitem{crankshaw2020inferline}
D.~Crankshaw, G.-E. Sela, X.~Mo, C.~Zumar, I.~Stoica, J.~Gonzalez, and
  A.~Tumanov, ``Inferline: latency-aware provisioning and scaling for
  prediction serving pipelines,'' in \emph{Proc. of the 11th ACM Symp. on Cloud
  Computing}, 2020, pp. 477--491.

\bibitem{nigade2022jellyfish}
V.~Nigade, P.~Bauszat, H.~Bal, and L.~Wang, ``Jellyfish: Timely inference
  serving for dynamic edge networks,'' in \emph{2022 IEEE Real-Time Systems
  Symp. (RTSS)}.\hskip 1em plus 0.5em minus 0.4em\relax IEEE, 2022, pp.
  277--290.

\bibitem{zeng2020distream}
X.~Zeng, B.~Fang, H.~Shen, and M.~Zhang, ``Distream: scaling live video
  analytics with workload-adaptive distributed edge intelligence,'' in
  \emph{Proc. of the 18th SenSys Conf.}, 2020, pp. 409--421.

\bibitem{Hou2023dystri}
X.~Hou, Y.~Guan, and T.~Han, ``{Dystri: A Dynamic Inference based Distributed
  DNN Service Framework on Edge},'' in \emph{ACM Int. Conf. Proceeding
  Series}.\hskip 1em plus 0.5em minus 0.4em\relax ACM, 2023, pp. 625--634.

\bibitem{Mendula2024deviceserversplit}
M.~Mendula, P.~Bellavista, M.~Levorato, and S.~L. {de Guevara Contreras},
  ``{Furcifer: a Context Adaptive Middleware for Real-world Object Detection
  Exploiting Local, Edge, and Split Computing in the Cloud Continuum},'' in
  \emph{2024 IEEE Int. Conf. on Pervasive Computing and Communications
  (PerCom)}, 2024, pp. 47--56.

\bibitem{Wu2022colocationinference}
J.~Wu, L.~Wang, Q.~Pei, X.~Cui, F.~Liu, and T.~Yang, ``{HiTDL: High-Throughput
  Deep Learning Inference at the Hybrid Mobile Edge},'' \emph{IEEE Transactions
  on Parallel and Distributed Systems}, vol.~33, no.~12, pp. 4499--4514, 2022.

\bibitem{hu2021rim}
Y.~Hu, W.~Pang, X.~Liu, R.~Ghosh, B.~Ko, W.-H. Lee, and R.~Govindan, ``Rim:
  Offloading inference to the edge,'' in \emph{Proc. of the Int. Conf. on
  Internet-of-Things Design and Implementation}, 2021, pp. 80--92.

\bibitem{Ananthanarayanan2020rocket}
\BIBentryALTinterwordspacing
``Microsoft rocket for live video analytics,'' pp. 1--6, 2020. [Online].
  Available:
  \url{https://www.microsoft.com/en-us/research/project/live-video-analytics/}
\BIBentrySTDinterwordspacing

\bibitem{Hung2018VideoEdge}
C.-C. Hung, G.~Ananthanarayanan, P.~Bodik, L.~Golubchik, M.~Yu, P.~Bahl, and
  M.~Philipose, ``Videoedge: Processing camera streams using hierarchical
  clusters,'' in \emph{2018 IEEE/ACM Symp. on Edge Computing (SEC)}, 2018, pp.
  115--131.

\bibitem{Cox2021interlayers}
B.~Cox, J.~Galjaard, A.~Ghiassi, R.~Birke, and L.~Y. Chen, ``{Masa: Responsive
  Multi-DNN Inference on the Edge},'' in \emph{2021 IEEE Int. Conf. on
  Pervasive Computing and Communications}, 2021, pp. 1--10.

\bibitem{raca2020beyond}
D.~Raca, D.~Leahy, C.~J. Sreenan, and J.~J. Quinlan, ``Beyond throughput, the
  next generation: A 5g dataset with channel and context metrics,'' in
  \emph{Procs. of the 11th ACM multimedia systems Conf.}, 2020, pp. 303--308.

\bibitem{Zhou2022trt}
Y.~Zhou and K.~Yang, ``{Exploring TensorRT to Improve Real-Time Inference for
  Deep Learning},'' in \emph{Proc. of HPCC}.\hskip 1em plus 0.5em minus
  0.4em\relax IEEE, 2022, pp. 2011--2018.

\bibitem{Demidovskij2020openvino}
A.~Demidovskij, A.~Tugaryov, A.~Suvorov, Y.~Tarkan, M.~Fatekhov, I.~Salnikov,
  A.~Kashchikhin, V.~Golubenko, G.~Dedyukhina, A.~Alborova, R.~Palmer,
  M.~Fedorov, and Y.~Gorbachev, ``{OpenVINO Deep Learning Workbench: A Platform
  for Model Optimization, Analysis and Deployment},'' in \emph{2020 IEEE 32nd
  Int. Conf. on Tools with Artificial Intelligence (ICTAI)}.\hskip 1em plus
  0.5em minus 0.4em\relax IEEE, 2020, pp. 661--668.

\bibitem{Kochura2020tpu}
Y.~Kochura, Y.~Gordienko, V.~Taran, N.~Gordienko, A.~Rokovyi, O.~Alienin, and
  S.~Stirenko, ``{Batch Size Influence on Performance of Graphic and Tensor
  Processing Units During Training and Inference Phases},'' in \emph{Proc. of
  ICCSEEA}, 2020, pp. 658--668.

\bibitem{Silfa2022rnn}
F.~Silfa, J.~M. Arnau, and A.~Gonz{\'{a}}lez, ``{E-BATCH: Energy-Efficient and
  High-Throughput RNN Batching},'' \emph{ACM Transactions on Architecture and
  Code Optimization}, vol.~19, no.~1, 2022.

\bibitem{zhang2017videostorm}
H.~Zhang, G.~Ananthanarayanan, P.~Bodik, M.~Philipose, P.~Bahl, and M.~J.
  Freedman, ``Live video analytics at scale with approximation and
  {Delay-Tolerance},'' in \emph{14th USENIX Symp. on Networked Systems Design
  and Implementation}, 2017, pp. 377--392.

\bibitem{Jiang2018chameleon}
J.~Jiang, G.~Ananthanarayanan, P.~Bodik, S.~Sen, and I.~Stoica, ``{Chameleon:
  scalable adaptation of video analytics},'' in \emph{Proc. of the 2018 Conf.
  of the ACM Special Interest Group on Data Communication}.\hskip 1em plus
  0.5em minus 0.4em\relax ACM, 2018, pp. 253--266.

\bibitem{Tan2020FastVA}
T.~Tan and G.~Cao, ``{FastVA: Deep Learning Video Analytics Through Edge
  Processing and NPU in Mobile},'' in \emph{IEEE Conf. on Computer
  Communications}.\hskip 1em plus 0.5em minus 0.4em\relax IEEE, 2020, pp.
  1947--1956.

\bibitem{Zhang2018awstream}
B.~Zhang, X.~Jin, S.~Ratnasamy, J.~Wawrzynek, and E.~A. Lee, ``{AWStream:
  adaptive wide-area streaming analytics},'' in \emph{Proc. of the 2018 Conf.
  of the ACM Special Interest Group on Data Communication}.\hskip 1em plus
  0.5em minus 0.4em\relax ACM, 2018, pp. 236--252.

\bibitem{Li2020reducto}
Y.~Li, A.~Padmanabhan, P.~Zhao, Y.~Wang, G.~H. Xu, and R.~Netravali,
  ``{Reducto: On-Camera Filtering for Resource-Efficient Real-Time Video
  Analytics},'' in \emph{Proc. of SIGCOMM}.\hskip 1em plus 0.5em minus
  0.4em\relax ACM, 2020, pp. 359--376.

\bibitem{Kang2017noscope}
D.~Kang, J.~Emmons, F.~Abuzaid, P.~Bailis, and M.~Zaharia, ``{NoScope:
  optimizing neural network queries over video at scale},'' \emph{Proc. of the
  VLDB Endowment}, vol.~10, no.~11, pp. 1586--1597, 2017.

\bibitem{Zhang2015vigil}
T.~Zhang, A.~Chowdhery, P.~V. Bahl, K.~Jamieson, and S.~Banerjee, ``{The Design
  and Implementation of a Wireless Video Surveillance System},'' in \emph{Proc.
  of the 21st Annual Int. Conf. on Mobile Computing and Networking}.\hskip 1em
  plus 0.5em minus 0.4em\relax ACM, 2015, pp. 426--438.

\bibitem{wang2022DRLTO}
J.~Wang, J.~Hu, G.~Min, W.~Zhan, A.~Y. Zomaya, and N.~Georgalas, ``Dependent
  task offloading for edge computing based on deep reinforcement learning,''
  \emph{IEEE Transactions on Computers}, vol.~71, no.~10, pp. 2449--2461, 2022.

\bibitem{liang2024splitstream}
Y.~Liang, S.~Zhang, and J.~Wu, ``Splitstream: Distributed and workload-adaptive
  video analytics at the edge,'' \emph{Journal of Network and Computer
  Applications}, vol. 225, p. 103866, 2024.

\bibitem{gao2024edgevision}
G.~Gao, Y.~Dong, R.~Wang \emph{et~al.}, ``Edgevision: Towards collaborative
  video analytics on distributed edges for performance maximization,''
  \emph{IEEE Transactions on Multimedia}, 2024.

\bibitem{Tan2021migserving}
C.~Tan, Z.~Li, J.~Zhang, Y.~Cao, S.~Qi, Z.~Liu, Y.~Zhu, and C.~Guo, ``{Serving
  DNN Models with Multi-Instance GPUs: A Case of the Reconfigurable Machine
  Scheduling Problem},'' \emph{arXiv}, 2021.

\bibitem{Gujarati2020Clockwork}
A.~Gujarati, R.~Karimi, S.~Alzayat, W.~Hao, A.~Kaufmann, Y.~Vigfusson, and
  J.~Mace, ``Serving {DNNs} like clockwork: Performance predictability from the
  bottom up,'' in \emph{14th USENIX Symp. on Operating Systems Design and
  Implementation (OSDI 20)}, 2020, pp. 443--462.

\bibitem{Zhang2023SHEPHERD}
H.~Zhang, Y.~Tang, A.~Khandelwal, and I.~Stoica, ``{SHEPHERD}: Serving {DNNs}
  in the wild,'' in \emph{20th USENIX Symp. on Networked Systems Design and
  Implementation}, 2023, pp. 787--808.

\end{thebibliography}

\end{document}